\documentclass[twocolumn,amsmath,amssymb,prd]{revtex4}

\def\al{\alpha}
\def\be{\beta}
\def\ga{\gamma}
\def\de{\delta}
\def\ep{\epsilon}

\def\et{\eta}
\def\th{\theta}

\def\ka{\kappa}
\def\la{\lambda}

\def\si{\sigma}

\def\ta{\tau}

\def\ph{\phi}

\def\ch{\chi}
\def\ps{\psi}
\def\om{\omega}

\def\De{\Delta}

\def\La{\Lambda}

\def\Om{\Omega}

\def\cl{{\cal L}}

\def\mn{{\mu\nu}}

\def\fr#1#2{{{#1}\over{#2}}}
\def\frac#1#2{{\textstyle{{#1}\over{#2}}}}
\def\half{{\textstyle{1\over 2}}}
\def\ol{\overline}
\def\prt{\partial}

\def\Re{\hbox{Re}\,}
\def\Im{\hbox{Im}\,}

\def\lsim{\mathrel{\rlap{\lower4pt\hbox{\hskip1pt$\sim$}}
    \raise1pt\hbox{$<$}}}
\def\gsim{\mathrel{\rlap{\lower4pt\hbox{\hskip1pt$\sim$}}
    \raise1pt\hbox{$>$}}}

\def\etal{{\it et al.}}

\def\vev#1{\langle {#1}\rangle}

\def\ket#1{|{#1}\rangle}

\def\sqr#1#2{{\vcenter{\vbox{\hrule height.#2pt
         \hbox{\vrule width.#2pt height#1pt \kern#1pt
         \vrule width.#2pt}
         \hrule height.#2pt}}}}

\newcommand{\beq}{\begin{equation}}
\newcommand{\eeq}{\end{equation}}
\newcommand{\bea}{\begin{eqnarray}}
\newcommand{\eea}{\end{eqnarray}}
\newcommand{\rf}[1]{(\ref{#1})}

\newcommand{\bM}{\begin{pmatrix}}
\newcommand{\eM}{\end{pmatrix}}

\def\psb{\ol\ps{}}

\def\mbf#1{\boldsymbol #1}

\def\syjm#1#2{{}_{#1}Y_{#2}}

\def\Q{\mathcal Q}

\def\K{\mathcal K}
\def\R{\mathcal R}

\def\pvec{\mbf p}

\def\sivec{\mbf\si}

\def\Svec{\mbf S}

\def\pmag{|\pvec|}

\def\punit{\hat p}

\def\epunit{\hat\ep}
\def\thunit{\hat\th}
\def\phunit{\hat\ph}

\def\phat{\mbf\punit}

\def\ephat{\mbf\epunit}
\def\thhat{\mbf\thunit}
\def\phhat{\mbf\phunit}

\def\Qhat{\widehat\Q}

\def\template#1#2#3#4{#1^{(#2)#4}_{#3}}
\def\acoef#1#2{\template{a}{#1}{#2}{}}

\def\ccoef#1#2{\template{c}{#1}{#2}{}}

\def\gzBcoef#1#2{\template{g}{#1}{#2}{(0B)}}
\def\goBcoef#1#2{\template{g}{#1}{#2}{(1B)}}
\def\goEcoef#1#2{\template{g}{#1}{#2}{(1E)}}
\def\HzBcoef#1#2{\template{H}{#1}{#2}{(0B)}}
\def\HoBcoef#1#2{\template{H}{#1}{#2}{(1B)}}
\def\HoEcoef#1#2{\template{H}{#1}{#2}{(1E)}}
\def\nr{{\rm NR}}
\def\nrtemplate#1#2#3{#1^{\nr#3}_{#2}}
\def\anr#1{\nrtemplate{a}{#1}{}}

\def\cnr#1{\nrtemplate{c}{#1}{}}

\def\gzBnr#1{\nrtemplate{g}{#1}{(0B)}}
\def\goBnr#1{\nrtemplate{g}{#1}{(1B)}}
\def\goEnr#1{\nrtemplate{g}{#1}{(1E)}}
\def\HzBnr#1{\nrtemplate{H}{#1}{(0B)}}
\def\HoBnr#1{\nrtemplate{H}{#1}{(1B)}}
\def\HoEnr#1{\nrtemplate{H}{#1}{(1E)}}

\def\ring#1{{\mathaccent'27 #1}}

\def\nrfctemplate#1#2{\nrtemplate{\ring{#1}}{#2}{}}
\def\anrfc#1{\nrfctemplate{a}{#1}}

\def\cnrfc#1{\nrfctemplate{c}{#1}}

\def\widecheck#1{\hskip#1pt\huge$\check{}$}
\def\bighacek#1#2{\vbox{\ialign{##\crcr\widecheck#2\crcr
  \noalign{\kern-9.5pt\nointerlineskip}
   $\hfil\displaystyle{#1}\hfil$\crcr}}}
\def\hb{\bighacek{b}{2}}
\def\hd{\bighacek{d}{2}}
\def\hg{\bighacek{g}{3}}
\def\hh{\bighacek{H}{5}{}}
\def\hH{\bighacek{H}{5}{}}
\def\hk{\bighacek{\K}{5}{}}

\def\mm{muonium}
\def\mmb{antimuonium}
\def\hm{${\rm H}_\mu$}
\def\hmb{$\overline{\rm H}_{\bar\mu}$}
\def\dm{${\rm D}_\mu$}
\def\tm{${\rm T}_\mu$}
\def\hetm{$^3{\rm He}^+_\mu$}
\def\hefm{$^4{\rm He}^+_\mu$}
\def\lixm{$^6{\rm Li}^{2+}_\mu$}
\def\linm{$^7{\rm Li}^{2+}_\mu$}
\def\bem{$^9{\rm Be}^{3+}_\mu$}
\def\bom{$^{11}{\rm B}^{4+}_\mu$}

\def\mr{m_{\rm r}}
\def\xm{\xi}

\def\codt{\cos{\om_\oplus T_\oplus}}
\def\sodt{\sin{\om_\oplus T_\oplus}}
\def\ctodt{\cos{2\om_\oplus T_\oplus}}
\def\stodt{\sin{2\om_\oplus T_\oplus}}
\def\ce{\cos\et}
\def\se{\sin\et}
\def\cc{\cos\ch}
\def\sc{\sin\ch}
\def\cto{\cos{\Om_\oplus T}}
\def\sto{\sin{\Om_\oplus T}}

\def\k{k}
\def\n{n}
\def\q{q}

\begin{document}
\title{Laboratory tests of Lorentz and CPT symmetry with muons} 

\author{Andr\'e H.\ Gomes,$^1$ 
V.\ Alan Kosteleck\'y,$^2$ 
and Arnaldo J.\ Vargas$^2$}

\affiliation{
$^1$Departamento de F\'\i sica,
Universidade Federal de Vi\c cosa,
36570-000 Vi\c cosa,
MG, Brazil\\
$^2$Physics Department, Indiana University, 
Bloomington, Indiana 47405, USA}

\date{IUHET 586, July 2014;
published as Phys.\ Rev.\ D {\bf 90}, 076009 (2014)}

\begin{abstract}

The prospects are explored for testing Lorentz and CPT symmetry 
in the muon sector
via the spectroscopy of muonium and various muonic atoms,
and via measurements of the anomalous magnetic moments 
of the muon and antimuon.
The effects of Lorentz-violating operators
of both renormalizable and nonrenormalizable dimensions
are included.
We derive observable signals,
extract first constraints from existing data
on a variety of coefficients for Lorentz and CPT violation,
and estimate sensitivities attainable in forthcoming experiments.
The potential of Lorentz violation to resolve 
the proton radius puzzle and the muon anomaly discrepancy
is discussed.

\end{abstract}

\maketitle

\section{Introduction}

Muons have played a significant role in testing relativity 
since their discovery in the 1930s
\cite{an}.
Indeed,
the first demonstration of time dilation
was the Rossi-Hall experiment
studying muons originating from cosmic rays
\cite{rh}.
As another example,
the clock hypothesis that 
acceleration {\it per se} has no affect on a clock's ticking rate
has been verified using muons in a ring accelerator
\cite{bailey}. 
 
In recent years,
the prospect of tiny deviations from relativity
has emerged as a promising candidate signal
for new physics coming from the Planck scale $M_P \simeq 10^{19}$ GeV,
following the demonstration that Lorentz invariance 
can naturally be broken 
in a unified framework of quantum gravity such as string theory
\cite{ksp}.
Driven by this prospect,
many high-precision tests of relativity
in different systems have been performed
to search for Lorentz violation
\cite{tables}.
Here,
we investigate the role of muons in this context,
focusing on 
laboratory studies using spectroscopy of muonic bound states
and measurements of the muon anomalous magnetic moment.

The general theoretical description of Lorentz violation
is provided by a realistic effective quantum field theory
called the Standard-Model Extension (SME)
\cite{ck,akgrav}.
The SME offers a theoretical framework for analysis 
based on General Relativity and the Standard Model (SM),
and it also characterizes CPT violation
\cite{ck,owg}.
Its Lagrange density consists of all coordinate-independent scalars 
built from the contractions of Lorentz-violating operators 
with coefficients determining the size of the associated effects.
The mass dimension of each coefficient
is fixed by the dimension $d$ of the operator,
with operators of larger $d$ often taken as higher-order terms 
in a low-energy expansion of the fundamental theory.
The restriction to terms containing 
operators of renormalizable dimensions $d\leq 4$
is called the minimal SME. 

The structure of the SME reveals that
Lorentz and CPT violation can be sector dependent,
with coefficients varying according to the particle species involved.
The size of Lorentz and CPT violation could conceivably increase with mass,
for example,
if the Yukawa-type couplings from spontaneous Lorentz violation
scale like the conventional Yukawa couplings in the SM.
Muon-sector experiments are of particular interest in this context
because they offer excellent prospects for a sensitive study 
of Lorentz and CPT violation in second-generation matter.
However,
given the extensive historical impact of research with muons 
and their comparatively widespread availability, 
surprisingly little is known about the SME muon sector
on both the theoretical and the experimental fronts.
For example,
inspection of the {\it Data Tables} 
\cite{tables}
reveals that existing constraints on Lorentz and CPT violation
involving muons comprise only a small fraction of the available limits.
The effects of minimal-SME coefficients
on the behavior of muons 
\cite{bkl}
have been studied at impressive sensitivities in the laboratory 
via muonium hyperfine spectroscopy
\cite{hughes01}
and via measurements of the anomalous magnetic moments
of the muon and antimuon 
\cite{bnl01,bnl08}.
The latter have also been used to place limits 
on nonminimal interaction terms with $d=5$
\cite{srf},
while minimal-SME interactions have been studied
in the context of muon decay
\cite{noordmans}.
A few constraints from astrophysical processes
have been obtained as well,
both for minimal-SME coefficients
\cite{ba07}
and for isotropic nonminimal operators with $d=5,6$
\cite{gm04,km13}.

In the present work,
we take advantage of the recently developed 
comprehensive approach to Lorentz-violating operators 
governing the propagation of a massive Dirac fermion
\cite{km13}
to study a broad range of effects involving
muon operators of both renormalizable and nonrenormalizable dimension.
Nonminimal terms produce effects that grow with energy,
so typically higher-energy experiments have greater sensitivity
to these effects.
However,
observable effects in the nonrelativistic limit
involve combinations of operators of arbitrary $d$,
and so studying these offers a different and powerful 
measure of sensitivity.
We therefore consider both nonrelativistic and relativistic experiments
in what follows.

For definiteness and simplicity,
the analysis in this work is restricted to Lorentz violation
in the muon sector.
Lorentz violation in other sectors
can better be sought with correspondingly dedicated experiments.
In the event that a nonzero signal is found,
comparison of results among different sectors
would be necessary to establish unequivocally the origin of the effect. 
This approach is justified because no compelling experimental evidence 
for Lorentz violation exists at present,
so the subject is currently in a search phase 
rather than a model-building phase.
Our analysis also focuses on effects originating from muon kinematics,
which provide the leading-order corrections from Lorentz violation
in the experiments we consider.
For example,
the leading-order correction in Coulomb gauge
is independent of the four-vector potential
in the bound states discussed below,
while the external magnetic fields used in all experiments 
are tiny compared to the muon mass
and hence their Lorentz-violating contributions are suppressed
by many orders of magnitude. 
Disregarding effects in interactions
also implies neglecting the Lorentz-violating flavor-changing operators 
that mix the charged leptons in the SME,
but as these necessarily also entail lepton-number violation 
they can plausibly be taken as suppressed 
relative to the effects we consider here. 

The analysis that follows can therefore be viewed as
primarily an investigation of the Lorentz- and CPT-violating 
kinetic Lagrange density for the muon and antimuon,
\beq
\cl \supset 
\half \psb (\ga^\mu i\prt_\mu - m_\mu + \Qhat) \ps 
+ {\rm h.c.}, 
\label{lag}
\eeq
where $\ps(x)$ is the muon quantum field,
$m_\mu$ is the muon mass,
and $\Qhat$ is a spinor-matrix operator 
describing all kinetic effects from Lorentz and CPT violation,
formed from derivatives $i\prt_\mu$
and SME coefficients for Lorentz and CPT violation.
By expanding $\Qhat$ in the basis of Dirac matrices in momentum space
and performing a decomposition in spherical coordinates,
the SME coefficients at each $d$ can be classified and enumerated
\cite{km13}. 
The freedom to perform field redefinitions in the theory
without changing the physics
implies that only certain combinations of these coefficients,
called effective coefficients,
are observable in a given experiment.
Under the assumptions made here, 
the muon sector with $d=3$
is found to have six independent effective coefficients
controlling CPT-even effects.
The sector with $d=4$ contains 30 independent effective coefficients,
of which 20 govern CPT-odd effects
and the other 10 are CPT even but observable
only if coordinate choices establishing the Minkowski metric
have otherwise been fixed. 
The nonminimal muon sector with $d=5$ has 65 independent effective coefficients
of which 20 are associated with CPT-odd operators,
while for $d=6$ we find 119 independent effective coefficients 
with 84 corresponding to CPT-odd operators.
Each observable effective coefficient represents a distinct physical way
to violate Lorentz symmetry.
As described in the sections below,
the muon experiments considered in this work
can access only a subset of these coefficients
in specific linear combinations,
but they nonetheless provide a broad-scope survey
of possible muon-sector effects 
and in many cases yield Planck-scale sensitivity.

Our investigations begin in Sec.\ \ref{Muonic bound states}
with the spectroscopy of muonic bound states.
Following some basics presented in Sec.\ \ref{Basics},
the effects of Lorentz and CPT violation
on muonium spectroscopy are considered in Sec.\ \ref{Muonium},
concerning first the hyperfine transitions
and then the $1S$-$2S$ transition and the Lamb shift.
We use existing data to place numerous first constraints
on nonrelativistic coefficients for Lorentz violation
and estimate possible sensitivities in some future experiments.
Section \ref{Muonic hydrogen} contains our discussion
of the spectroscopy of muonic atoms,
with a focus on muonic hydrogen.
After some general considerations,
the prospects are investigated for future searches 
using sidereal variations in Zeeman transitions.
In Sec.\ \ref{Proton radius puzzle}
we address the proton radius puzzle 
in the context of Lorentz violation,
outlining the requirements for a resolution
and the ensuing predictions for future experiments.
Section \ref{Negligible magnetic field}
describes a scheme for performing searches for Lorentz violation
when the applied magnetic field is effectively negligible,
as is the case in current experiments. 
The spectroscopy of various other muonic atoms,
including among others muonic deuterium and muonic helium,
is discussed in Sec. \ref{Other muonic atoms}.

Section \ref{Muon magnetic moment}
focuses on measurements of the muon and antimuon magnetic moments.
Some relevant theory is presented in Sec.\ \ref{Theory}.
We then turn to comparisons of the muon and antimuon
in Sec.\ \ref{Muon-antimuon comparison},
where techniques for extracting constraints 
on CPT-odd and CPT-even operators are described 
and used in conjunction with existing data
to place numerous first bounds on nonminimal coefficients
for Lorentz and CPT violation. 
Another potential signal is sidereal variations,
which are the subject of Sec.\ \ref{Sidereal variations}
and also lead to a variety of first bounds.
In Sec.\ \ref{Annual variations} we consider the potential 
of future analyses to incorporate signals involving annual variations,
some of which are tied to the Earth's changing boost
as it orbits the Sun.
Existing measurements of the muon anomaly lack full concordance 
with SM calculations,
and in Sec.\ \ref{The anomaly discrepancy}
we consider the prospects of accounting for the anomaly discrepancy
using Lorentz violation and describe some predicted signals
in future experiments.
Finally,
Sec.\ \ref{Summary} concludes with a summary
and discussion of other possibilities for future exploration
of muon-sector Lorentz violation.

With a few exceptions described in the text,
the notation and conventions throughout this work 
are those of Ref.\ \cite{km13}.
For simplicity,
the index $\mu$ indicating the muon sector
is omitted from all coefficients.

\section{Muonic bound states}
\label{Muonic bound states}

This section focuses on searches for Lorentz and CPT violation
using spectroscopy of exotic bound states 
having a muon or antimuon as a constituent.
Among the many possible systems are
various onia, which are bound states of a muon with another lepton
of opposite charge;
muonic atoms or ions, 
which are atoms or ions with an electron replaced by a muon;
and hadron-muon bound states.
Recent high-precision spectroscopy has been performed with 
muonium 
\cite{hughes01},
which is the bound state of 
an antimuon $\mu^+$ and an electron $e^-$,
and with muonic hydrogen \hm\
\cite{antognini},
which is the bound state of a proton $p$ with a muon $\mu^-$.
Spectroscopy of muonic deuterium \dm\
and of the muonic-helium ions \hetm\ and \hefm\
is in the offing
\cite{deuterium,crema}.
Future studies of other exotic bound states involving muons
such as muonic tritium \tm\
\cite{fujiwara}
or various muonic ions
such as \lixm, \linm, \bem, or \bom\
\cite{drakebyer}
may be of interest as well.
Here,
we consider the effects of Lorentz and CPT violation
on the spectroscopy of all these muonic bound states.

With minor notation and interpretational changes,
many of the results that follow
can be directly transcribed to many other muonic bound states
of potential interest.
One exception is true muonium,
the bound state of a muon with an antimuon,
which may be observed and studied in electron-positron colliders
\cite{bl}
or in fixed-target electroproduction
\cite{hps}. 
The transcription in this case requires taking into account
the equal masses of the two constituents 
and the Lorentz-violating corrections for both the muon and the antimuon.
With specific sign changes as indicated in the text below,
our results are also directly applicable to antimuonium,
which is a bound state of a muon $\mu^-$ and a positron $e^+$,
and to antimuonic antihydrogen \hmb, 
which is a bound state of an antiproton $\overline p$ and an antimuon $\mu^+$.
Should precision spectroscopy of these exotic antiatoms 
eventually become feasible,
direct CPT tests comparing \mm\ with \mmb\ 
and \hm\ with \hmb\ could be performed.
However,
with current technology the search for CPT violation
in the \mm\ and \hm\ systems is of necessity reliant on
either studying sidereal variations
or comparing observed transition frequencies with theoretical calculations.

\subsection{Basics}
\label{Basics}

The bound states of interest here 
involve two particles of different masses,
one of which may be an atomic nucleus.
The rotational symmetry of the conventional interactions
ensures conservation of total angular momentum $\mbf F$ of the system
and implies that energy levels 
labeled with the corresponding quantum number $F$
are $(2F+1)$-fold degenerate. 
The asymmetry of the masses leads to a hierarchy
in the angular-momentum couplings,
which causes the hyperfine structure to be smaller than the fine structure
by the ratio of the lighter to the heavier mass.
In the absence of Lorentz violation,
the general features of the spectra of the muonic bound states of interest
therefore largely parallel those of hydrogen,
with appropriate scalings originating in the mass, charge, 
and nuclear-spin differences.
For example, 
the ground-state energy of \hm\
is larger than that of H 
by a factor of the ratio of the corresponding reduced masses,
which is about 186.
One notable exception is the Lamb shift in \hm,
which is enhanced by a factor of order $1/\al^2$
via radiative corrections in quantum field theory,
producing a 2S level lying well below the 2P level
\cite{jaw,eides}.

Some spectroscopic experiments of interest
are performed with the system placed in a magnetic field,
which breaks rotational symmetry and hence also conservation of $\mbf F$,
thereby lifting the $(2F+1)$-fold degeneracies of the energy levels.
In this work,
we treat the applied magnetic field as uniform and constant.
We also assume the induced level shifts 
are smaller than the fine structure,
although possibly smaller or larger than the hyperfine structure
so that both the hyperfine Zeeman and the hyperfine Paschen-Back limits
can be considered.
In this scenario
the magnitude of the total angular momentum $\mbf J$ 
of the lighter particle,
which is the sum $\mbf J=\mbf L + \mbf S$
of its orbital and spin angular momenta,
can be approximated as independently conserved.
The corresponding quantum number $J$ can therefore
be used to label states even when $F$ cannot.

Since combinations of Lorentz boosts generate rotations,
violations of Lorentz symmetry are accompanied
by violations of rotation invariance in generic observer frames.
The presence of Lorentz violation 
can therefore lift some or all of the $(2F+1)$-fold degeneracies 
in the energy levels of the free system,
and it can modify the level splittings arising from an applied magnetic field.
Unless otherwise specified,
in this work
we assume the effects from Lorentz and CPT violation are small
compared to those from any magnetic field present.
The lifting of the degeneracies by the magnetic field
then has the technical advantage of 
avoiding degenerate perturbation theory 
in calculations of Lorentz-violating corrections.
For consistency,
we also assume that the Lorentz violation is sufficiently small
to ensure maintenance of the perturbative regime 
where stability and causality are preserved in concordant frames
\cite{akrl}. 

The muon is nonrelativistic in all the bound systems considered here,
so for small Lorentz and CPT violation in the muon sector 
the dominant perturbations to the spectra 
arise from the nonrelativistic limit.
In the Coulomb gauge,
all relevant contributions from the electromagnetic interactions
arise from the zero component of the covariant derivative
acting on the muon field 
or equivalently in momentum space from the canonical energy of the muon,
which in the nonrelativistic limit reduces to the muon mass.
The leading-order perturbation is therefore independent
of the electromagnetic potential,
so it suffices for calculational purposes
to consider only the Lorentz-violating corrections 
to the nonrelativistic free motion of the muon.
This is physically plausible
because the binding energy of the system is small compared to the muon mass,
and it also matches established results 
for related analyses of Lorentz violation in conventional atoms
\cite{kla}.

A complete classification of Lorentz-violating terms 
of arbitrary mass dimension
that can appear in the quadratic Lagrange density for a massive Dirac fermion
is given in Ref.\ \cite{km13},
along with a derivation of the corresponding nonrelativistic hamiltonian.
To apply this framework in the present context,
we can work in the zero-momentum frame of the two-particle atom,
which in typical applications can be taken as the laboratory frame.
The leading-order corrections due to Lorentz and CPT violation
for a nonrelativistic muon of momentum $\pvec$
are then described by an effective hamiltonian $\de h^\nr (\pvec)$ 
that can be split into four types of terms,
according to whether the physics is 
spin independent or dependent
and whether the CPT effects are even or odd.

For experimental applications,
it is convenient to decompose $\de h^\nr$ in spherical coordinates
because sensitivity to rotational symmetry 
is the key to many searches for Lorentz violation.
Given the unit momentum vector $\phat = \pvec/\pmag$,
we can define spherical polar angles $\th$, $\ph$ in momentum space
by $\phat = (\sin\th\cos\ph,\sin\th\sin\ph,\cos\th)$.
A basis of unit vectors can be chosen as
$\ephat_r = \phat$, $\ephat_\pm = (\thhat \pm i\phhat)/\sqrt{2}$,
where $\thhat$ and $\phhat$ are the standard unit vectors
associated with the polar angle $\th$ and azimuthal angle $\ph$.
The result of decomposing $\de h^\nr$ can then be expressed as 
\cite{km13}
\beq
\de h^\nr
=h_0+h_r\sivec\cdot\ephat_r
+h_+\sivec\cdot\ephat_-
+h_-\sivec\cdot\ephat_+,
\label{FEH}
\eeq
where $\sivec = (\si^1, \si^2, \si^3)$ 
contains the three Pauli matrices.

The expression \rf{FEH} contains four component hamiltonians
$h_0$, $h_r$, $h_\pm$
that depend on the magnitude and direction
of the momentum and on SME coefficients for Lorentz and CPT violation.
The spin-independent component $h_0$
can be written as 
\beq
h_0 =
\sum_{kjm} \pmag^k 
~\syjm{0}{jm}(\phat) 
\left( \anr{\k jm}-\cnr{\k jm}\right),
\label{FEHSI}
\eeq
while the spin-dependent terms take the form
\bea
h_r &=&
\sum_{kjm} \pmag^k 
~\syjm{0}{jm}(\phat) 
\left( -\gzBnr{\k jm} +\HzBnr{\k jm} \right),
\nonumber \\
h_{\pm} &=&
\sum_{\k jm} \pmag^\k 
~\syjm{\pm 1}{jm}(\phat) 
\bigg[i\goEnr{\k jm}-i\HoEnr{\k jm}
\nonumber \\ 
&& 
\hskip 70pt
\pm\left(\goBnr{\k jm}-\HoBnr{\k jm}\right)
\bigg].
\qquad
\label{FEHSD}
\eea
Here,
the eight sets of quantities 
$\anr{\k jm}$, 
$\cnr{\k jm}$,
$\gzBnr{\k jm}$, 
$\goBnr{\k jm}$,
$\goEnr{\k jm}$,
$\HzBnr{\k jm}$,
$\HoBnr{\k jm}$,
$\HoEnr{\k jm}$
are nonrelativistic effective coefficients
for Lorentz and CPT violation.
The relationships between these nonrelativistic coefficients
and the complete set of spherical coefficients 
governing Lorentz and CPT violation in the muon sector
are given in Eqs.\ (111) and (112) of Ref.\ \cite{km13}.
For the antimuon,
the signs of the $a$- and $g$-type coefficients
in $\de h^\nr$ are reversed.
The allowed ranges of indices on all the coefficients
and their counting for given $k$ are summarized
in Table IV of Ref.\ \cite{km13}.
To avoid potential confusion with the principal quantum number $n$
of the exotic atoms considered here,
we use the notation $k$
instead of $n$ for the first index on these coefficients.
Also, 
in the above equations
the quantities $\syjm{s}{jm}(\phat)\equiv\syjm{s}{jm}(\th,\ph)$
are spin-weighted spherical harmonics of spin weight $s$,
with the usual spherical harmonics arising for $s=0$ as 
$\syjm{0}{jm}(\phat)\equiv Y_{jm}(\th,\ph)$.
Some key features of spin-weighted spherical harmonics
can be found in Appendix A of Ref.\ \cite{km09}.
Note that the indices $j$, $m$ characterize the rotational properties 
of the spherical harmonics
and thereby of the associated operators for Lorentz violation.
The reader is cautioned that these indices are distinct 
from the angular-momentum quantum numbers $J$, $M$ 
of a muonic bound state.

To determine the dominant level shifts 
arising from Lorentz and CPT violation
requires calculation of the expectation values of $\de h^\nr$
in the unperturbed eigenstates of the exotic atom,
which are the Schr\"odinger-Coulomb eigenfunctions
for the reduced mass $\mr$.
Inspection reveals that many of these expectation values vanish.
The angular-momentum wave functions are parity eigenstates,
so the expectation values of odd-parity Lorentz-violating operators are zero. 
This implies that only coefficients with even $\k\equiv 2q$ can contribute.
Also,
in the presence of a magnetic field,
only components of the Lorentz-violating operators
projected in the direction of the field play a role 
because the azimuthal pieces average to zero.
Choosing laboratory coordinates with the field along the $z$ direction,
this implies that only coefficients with $m=0$ are relevant.
Moreover,
the $E$-type coefficients also fail to contribute
because the corresponding operators are proportional
to $\syjm{+1}{j0}(\phat)\sivec\cdot\phhat$,
which precesses about the magnetic field.
These results and other more specific ones described below 
significantly reduce the calculations required
to obtain the dominant level perturbations.

Given a nonzero expectation value of $\de h^\nr$,
evaluation of the part involving a power $\pmag^k$ 
of the momentum magnitude can be performed directly 
because it is independent of the angular-momentum couplings.
For $k=0$, 2, and 4 we obtain
\bea
\vev{\pmag^0}_{nL}&=& 1,
\nonumber\\
\vev{\pmag^2}_{nL}&=& \left(\dfrac{\al \mr}{n}\right)^2,
\nonumber\\
\vev{\pmag^4}_{nL}&=&
\left(\dfrac{\al \mr}{n}\right)^4
\left(\dfrac{8n}{2L+1}-3\right),
\label{radialExp}
\eea
where $\al$ is the fine-structure constant,
$\n$ is the principal quantum number,
and $L$ is the orbital angular-momentum quantum number. 
In this work,
we disregard operators $\pmag^k$ with $k>4$,
which have expectation values diverging
for states with small values of $n$ and $L$.
Although this technical issue can in principle be avoided by regularizing,
the physical scale of the expectation values
is set by the factor $(\al \mr)^k$,
which for $k>4$ typically introduces 
only unobservable corrections 
to the transition frequencies.
For example,
even comparatively large coefficients for Lorentz violation
$\anr{6j0} \simeq 1$ GeV$^{-5}$
would lead only to frequency shifts
of order $10^{-9}$ Hz in \mm\ and $10^{5}$ Hz in \hm,
due to the appearance of the factor $(\al \mr)^6$.

The calculation of expectation values
yields the shifts in the transition frequencies.
The sizes of these shifts are set by the coefficients for Lorentz violation,
which in the above expression for $\de h^\nr$ 
are defined in the zero-momentum frame of the two-particle atom.
However,
this frame is noninertial due to the rotation of the Earth about its axis 
and, to a lesser extent,
due to the revolution of the Earth about the Sun.
A reasonable approximation to an inertial frame
on the time scale of experiments
is the canonical Sun-centered frame
\cite{sunframe,tables}
widely used to report results of searches for Lorentz and CPT violation,
which has $Z$ axis aligned with the Earth's rotation axis,
$X$ axis pointing from the Sun towards the vernal equinox,
and time $T$ with origin 
chosen by convention at the vernal equinox 2000.
In this frame,
the coefficients for Lorentz violation can be approximated as constants, 
so the rotation of the Earth introduces
time dependence in some laboratory-frame SME coefficients 
and hence sidereal variations in physical observables 
\cite{ak98}.
The spherical decomposition greatly simplifies
the calculation of these variations
because the two frames are related by a rotation.
Indeed,
the sidereal dependence of the transition frequencies
induced by a particular coefficient 
is essentially determined by its azimuthal index $m$. 
The general expression relating laboratory-frame coefficients
to those in the Sun-centered frame 
is given by Eq.\ (139) of Ref.\ \cite{km09}.
Results specific to the experiments considered here
are presented in the subsections that follow.

\subsection{Muonium}
\label{Muonium}

In this subsection,
we consider the effects of Lorentz and CPT violation on \mm\ spectroscopy.
The full perturbation hamiltonian $\de h^\nr$ is used
to determine the shifts in $1S$ hyperfine transitions, 
while the shift in the $1S$-$2S$ transition 
and the Lamb shift are calculated using
the spin-independent perturbation.
Existing experimental data are used to place
first constraints on some coefficients for Lorentz violation.

\subsubsection{Hyperfine transitions}
\label{Hyperfine transitions}

In a magnetic field
the ground state of \mm\ splits into four sublevels,
labeled 1, 2, 3, 4 
in order of decreasing energy.
Precision spectroscopy of the \mm\ $1S$ hyperfine transitions
$\nu_{12}$ and $\nu_{34}$ 
has been performed in a comparatively strong magnetic field
of about 1.7 T
\cite{liu99,hughes01}.
In this setup,
the four levels can be labeled as $\ket{m_S, m_I}$,
where $S$ and $I$ are the electron and muon
spin quantum numbers, respectively. 
The frequency $\nu_{12}$ corresponds to the transition
$\ket{1/2,1/2} \leftrightarrow \ket{1/2,-1/2}$,
while $\nu_{34}$ corresponds to 
$\ket{-1/2,-1/2} \leftrightarrow \ket{-1/2,1/2}$.

The dominant shifts $\de\nu_{12}$ and $\de\nu_{34}$
induced by Lorentz violation in these transition frequencies 
are given by the expectation values
of the part $\de h^\nr_{nS}$ 
of the perturbation hamiltonian \rf{FEH}
for antimuons
that affects the $n S_{1/2}$ levels,
\bea
\de h^\nr_{nS}&=& 
\sum_{q=0}^2\pmag^{2q} 
\bigg[
-(\anrfc{2q} +\cnrfc{2q})
\nonumber\\
&& 
\hskip 5pt
+\left( \gzBnr{(2q)10} + \HzBnr{(2q)10} \right) 
Y_{10}(\phat) \sivec\cdot \phat
\nonumber\\ 
&& 
\hskip 5pt
-\sqrt{2} \left( \goBnr{(2q)10} + \HoBnr{(2q)10} \right) 
\syjm{1}{10}(\phat) \sivec\cdot\thhat\bigg],
\hskip 15pt
\label{1S} 
\eea
where $\anrfc{2q} \equiv \anr{(2q)00}/\sqrt{4\pi}$ and
$\cnrfc{2q} \equiv \cnr{(2q)00}/\sqrt{4\pi}$
are isotropic nonrelativistic coefficients
\cite{km13}. 
This expression contains only nonrelativistic coefficients with $j\le 1$
because the expectation values of operators
associated with other coefficients vanish.
To illustrate this,
suppose $T_{jm}$ is a spherical-tensor operator 
and $\ket{j'm'}$ is an angular-momentum eigenstate.  
The Wigner-Eckart theorem 
\cite{we}
then implies that the expectation value 
$\vev{ j' m'|T_{jm}|j' m'}$ vanishes if $2j'<j$. 
In the present case the angular momentum of each fermion is $1/2$,
so only coefficients with $j\le 1$ contribute.  

The relevant eigenstates 
$\ket{m_S, m_I}$
for the perturbative calculation are the products
of the Schr\"odinger-Coulomb ground state 
and the appropriate Pauli spinors.
The comparatively strong magnetic field
ensures that the transitions are essentially muon-spin flips,
so any effects from Lorentz and CPT violation
in the electron sector can reasonably be disregarded.
The nonzero expectation values with respect to $\ket{m_S,m_I}$ are given by
\bea
\vev{Y_{10}(\phat)\sivec\cdot\phat }&=&
\dfrac{1}{\sqrt{3\pi}} m_I,
\nonumber \\
\vev{\syjm{1}{10}(\phat)\sivec\cdot\thhat }&=& 
-\sqrt{\dfrac{2}{3 \pi}} m_I.
\label{mSmI}
\eea
Incorporating also the results \rf{radialExp}
reveals that the frequency shifts take the form 
\bea
\de\nu_{12}&=& -\de \nu_{34}
\nonumber \\
&&
\hskip-30pt
= \sum_{q=0}^2 \dfrac{1}{\sqrt{12\pi^3}}
(\al \mr)^{2q}(1+4\de_{q2})
\nonumber\\ 
&& 
\hskip-10pt
\times \left( \gzBnr{(2q)10} +\HzBnr{(2q)10}
+2 \goBnr{(2q)10} +2\HoBnr{(2q)10} \right),
\nonumber \\
\label{Mhyper}
\eea
where $\de_{q2}=1$ when $q=2$ and vanishes otherwise, as usual.
Note that the condition 
$ \de\nu_{12}+\de \nu_{34}=0$ 
is a specific prediction of the present theoretical framework.
Note also that the corresponding frequency shifts for \mmb\
are given by changing the signs of the $g$-type coefficients
in this result.
If it should become practical to perform \mmb\ spectroscopy,
then a direct comparison of the hyperfine frequency shifts
for \mm\ and \mmb\ would make possible 
independent measurements of the $g$-type coefficients. 

The result \rf{Mhyper} extends the previous expression 
reported in Ref.\ \cite{bkl},
$\de\nu_{12}= -\de \nu_{34} = -\widetilde b_3^*/\pi$,
which involves the coefficient combination
\cite{tables}
$\widetilde b_3^* = b_3 + m_\mu d_{30} + H_{12}$
associated with certain minimal-SME operators of mass dimensions 3 and 4.
The present result includes also contributions 
from the minimal-SME coefficients $g_{\la\mu\nu}$,
along with effects from many operators of nonminimal mass dimensions.
In the limit where all nonminimal coefficients
are set to zero,
the minimal-SME result is recovered via the identity
\beq 
\gzBnr{010} + 2\goBnr{010} + \HzBnr{010} + 2\HoBnr{010} 
\supset
\sqrt{12\pi}~\widetilde{b}_3^*,
\eeq
where 
$\widetilde{b}_3^*$ is now defined to contain contributions
from $g_{\la\mu\nu}$ as well.

The connection between the coefficients in the Sun-centered frame
and those in the laboratory frame takes the generic form 
\bea
\K_{\k 10}^{\rm lab}&=&
\K_{\k 10}^{\rm Sun} 
\cos\chi
-{\sqrt{2}}~ 
\Re\K_{\k 11}^{\rm Sun}~
{\sin\chi}
\cos\om_\oplus T
\nonumber\\
&&
+{\sqrt{2}}~ 
\Im\K_{\k 11}^{\rm Sun}~ 
{\sin\chi}
\sin\om_\oplus T, 
\label{rot}
\eea
where $\ch$ is 
the angle between the laboratory magnetic field
and the Earth's rotational axis
and $\om_\oplus \simeq 2\pi/(23 {\rm ~h} ~56 {\rm ~m})$
is the Earth's sidereal frequency.
This explicitly displays the time variations 
in the laboratory-frame coefficients for Lorentz violation
induced by the rotation of the Earth.
In general,
the variations of a coefficient with index $m$ 
occur at the $m$th harmonic of $\om_\oplus$,
but in the present context the only relevant frequency 
is $\om_\oplus$ itself because
only coefficients with $j\leq 1$ play a role.
In general,
other frequencies appear that are associated with the revolution
of the Earth around the Sun,
but the corresponding effects 
are suppressed by a factor of the Earth's orbital speed 
$\be_\oplus\simeq 10^{-4}$.
A detailed analysis of these effects is possible in principle
but lies outside our present scope.
The treatment would follow a path analogous to that taken 
in Sec.\ \ref{Annual variations} below, 
which investigates annual variations in experiments 
on the anomalous magnetic moments of the muon and antimuon.
The attainable sensitivities via \mm\ hyperfine spectroscopy
would be some orders of magnitude weaker
but would involve different linear combinations 
of coefficients for Lorentz violation.  

The published experimental analysis constraining the sidereal variations 
of $\nu_{12}$ and $\nu_{34}$ from existing data
\cite{hughes01}
can be combined with the above results 
to extract constraints on the coefficients for Lorentz violation.
Hughes \etal\ found the data contained
no sidereal variation to $\pm$20 Hz at one standard deviation.
Here,
we adopt a limit on time variations of $\de\nu_{12}$ 
at the sidereal frequency $\om_\oplus$
corresponding to no signal within $\pm$32 Hz
at the 90\% confidence level.
Noting that $\ch\simeq 90^\circ$ in this experiment 
yields the bound 
\bea
&&
\hskip-20pt
\bigg[\sum_{m\in\{1,-1\}}
\bigg| \sum_{q=0}^{2}(\al \mr)^{2q}(1+4\de_{2q})
\bigg(\gzBnr{(2q)1m}
+\HzBnr{(2q)1m}
\nonumber\\ 
&& 
\hskip80pt
+2\goBnr{(2q)1m}
+2\HoBnr{(2q)1m}\bigg)
\bigg|^2 \bigg]^{1/2}
\nonumber\\
&<&
\dfrac{\sqrt{3\pi^3}}{|\sin{\chi}|} (32 {\rm ~Hz})
\simeq 2 \times 10^{-22} {~\rm GeV}.
\label{NRSDB}
\eea
in the Sun-centered frame.

Insight into the reach of this bound 
can be obtained by extracting from it
the attained sensitivities to individual coefficients 
under the assumption that only one coefficient is nonzero at a time.
These sensitivities are useful for several purposes including,
for example,
for comparisons between experiments and for constraining models.
Using this assumption,
Table \ref{table: NRHyper} 
provides estimated sensitivities in the Sun-centered frame 
to nonrelativistic coefficients with $k\leq 4$ 
that contribute to sidereal effects in \mm\ hyperfine splittings.
Several of these results represent first constraints
in the literature.

\renewcommand\arraystretch{1.5}
\begin{table}
\caption{Constraints on the moduli of the real and imaginary parts 
of muon nonrelativistic coefficients 
determined from \mm\ hyperfine transitions.}
\setlength{\tabcolsep}{7pt}
\begin{tabular}{cl}
\hline
\hline
Coefficient & \qquad Constraint on 
\\
[-4pt]
$\K$ & \qquad $|\Re\K|$, $|\Im\K|$ 
\\
\hline
$\HzBnr{011}$,		$\gzBnr{011}$	&	\quad	$<2 \times10^{-22}\,$GeV	\\
$\HoBnr{011}$,		$\goBnr{011}$	&	\quad	$<7\times10^{-23}\,$GeV	\\
$\HzBnr{211}$,		$\gzBnr{211}$	&	\quad	$<1 \times10^{-11}\,\text{GeV}^{-1}$	\\
$\HoBnr{211}$,		$\goBnr{211}$	&	\quad	$<6 \times10^{-12}\,\text{GeV}^{-1}$	\\
$\HzBnr{411}$,		$\gzBnr{411}$	&	\quad	$<2 \times10^{-1}\,\text{GeV}^{-3}$	\\
$\HoBnr{411}$,		$\goBnr{411}$	&	\quad	$<8 \times10^{-2}\,\text{GeV}^{-3}$	\\
\hline\hline
\end{tabular}
\label{table: NRHyper}
\end{table}

Since each nonrelativistic coefficient
is a linear combination of spherical coefficients
of different mass dimensions
\cite{km13},
the bound \rf{NRSDB} also can be interpreted
in terms of constraints on spherical coefficients.
As a simple illustration of this connection,
consider the nonrelativistic coefficient $\gzBnr{010}$
and suppose that only the spherical coefficients 
$\gzBcoef{d}{njm}$ with $n=0$, $j=1$, $m=0$ are nonzero.
Then,
$\gzBnr{010}=\sum_d m_\mu^{d-3} \gzBcoef{d}{010}$ 
is an infinite sum of spherical coefficients
of arbitrary even $d\geq 4$.
This confirms that nonrelativistic experiments
can access nonminimal coefficients of arbitrary dimensionality.
Table \ref{table: Hyper} 
collects the corresponding estimated sensitivities 
in the Sun-centered frame
to spherical coefficients with $d\leq 8$
under the assumption of only one nonzero coefficient at a time,
as before.

\renewcommand\arraystretch{1.5}
\begin{table}
\caption{Constraints on the moduli of the real and imaginary parts 
of muon spherical coefficients determined from \mm\ hyperfine transitions.}
\setlength{\tabcolsep}{7pt}
\begin{tabular}{cl}
\hline\hline
Coefficient & \qquad Constraint on 
\\
[-4pt]
$\K$ & \qquad $|\Re\K|$, $|\Im\K|$ 
\\
\hline
$ H_{011}^{(3)(0B)}$	&	\quad	$<5\times10^{-23}$ GeV	\\
$ g_{011}^{(4)(0B)}$	&	\quad	$<5\times10^{-22}$	\\
$ H_{011}^{(5)(0B)}$	&	\quad	$<5\times10^{-21}$ GeV$^{-1}$	\\
$ g_{011}^{(6)(0B)}$	&	\quad	$<5\times10^{-20}$ GeV$^{-2}$	\\
$ H_{011}^{(7)(0B)}$	&	\quad	$<4\times10^{-19}$ GeV$^{-3}$	\\
$ g_{011}^{(8)(0B)}$	&	\quad	$<4\times10^{-18}$ GeV$^{-4}$	\\
\hline
\hline
\end{tabular}
\label{table: Hyper}
\end{table}

Future experiments are likely to improve on the results 
listed in Tables \ref{table: NRHyper} and \ref{table: Hyper}.
For example,
the proposed \mm\ Hyperfine Structure (MuHFS) experiment 
\cite{shimomura}
at J-PARC
would be capable of measuring 
the \mm\ ground-state hyperfine splitting
to a few ppb.
This would lead to improvements of about a factor of five 
over the values listed in the above tables.

\subsubsection{1$S$-2$S$ transition and Lamb shift}
\label{1S-2S transition and Lamb shift}

The shifts \rf{Mhyper} in the hyperfine transition frequencies
are independent of the isotropic coefficients
$\anrfc{2q}$ and $\cnrfc{2q}$
appearing in the perturbation hamiltonian $\de h^\nr_{nS}$
given in Eq.\ \rf{1S}.
This is unsurprising
because the hyperfine transitions involve spin flips,
while $\anrfc{2q}$ and $\cnrfc{2q}$
control spin-independent contributions to the hamiltonian.
Indeed,
only transitions with $\De L\ne 0$ or $\De n\ne 0$ 
acquire shifts depending on these isotropic coefficients.

One transition of this kind that can be experimentally studied in \mm\
is the $1S_{1/2}$-$2S_{1/2}$ transition.
The lesser attainable measurement precision 
of the corresponding frequency $\nu_{1S2S}$ 
compared to the studies of hyperfine transitions 
implies that it is reasonable to disregard
spin-dependent terms in calculating 
the Lorentz-violating shift $\de\nu_{1S2S}$. 
As before,
we neglect possible contributions from 
Lorentz violation in the electron sector,
which in any case can be investigated at higher precision 
using other systems such as hydrogen
\cite{hydrogen}.

Taking the appropriate expectation values of Eq.\ \rf{1S}
and applying the result \rf{radialExp} yields the frequency shift 
for the $1S_{1/2}$-$2S_{1/2}$ transition as
\beq
\de \nu_{1S2S}=
\dfrac{3(\mr \al)^2}{8\pi}
[\anrfc{2}+ \cnrfc{2} 
+\frac{67}{12} (\mr \al)^2 
\left( \anrfc{4}+ \cnrfc{4}\right)].
\label{1s2s}
\eeq
This result is independent of the presence of a magnetic field 
and also of the hyperfine sublevel involved in the transition.
It represents a rotationally invariant 
but Lorentz- and CPT-violating shift in the transition frequency.
Note that the corresponding shift for \mmb\ is given by 
changing the signs of the coefficients
$\anrfc{2}$ and $\anrfc{4}$,
so comparisons of \mm\ and \mmb\ would permit
direct measurement of the isotropic $a$-type coefficients.

The rotational invariance of the shift \rf{1s2s}
ensures that sidereal variations of $\de \nu_{1S2S}$
of frequency $\om_\oplus$ are absent. 
Also,
although the Lorentz violation implies
an annual variation in $\de \nu_{1S2S}$
induced by the revolution of the Earth about the Sun,
this variation is suppressed by the orbital speed 
$\be_\oplus\simeq10^{-4}$  
and hence attainable constraints are of only limited interest. 
However,
the shift $\de \nu_{1S2S}$ does represent
a predicted physical effect.

One way to estimate a bound on this effect 
is to compare the observed experimental value 
$\nu_{1S2S}^{\rm expt}$ 
with the theoretical value 
$\nu_{1S2S}^{\rm th}$ 
calculated in conventional quantum electrodynamics,
requiring that the Lorentz-violating contribution
be no larger than the difference between them.
For illustrative purposes,
we adopt the experimental value 
\cite{meyer}
$\nu_{1S2S}^{\rm expt}=2 455 528 941.0(9.8)$ MHz 
and the theoretical value
\cite{eides} 
$\nu_{1S2S}^{\rm th}=2 455 528 935.7 (0.3)$ MHz.
Some care is required in using the latter
as it depends partly on other experimental measurements,
including the Rydberg constant, 
the fine-structure constant,
and the muon-electron mass ratio.  
However,
the first two of these are determined by non-muonic experiments
\cite{mohr},
so they can reasonably be treated as independent
of the nonrelativistic coefficients in the muon sector.
The determination of the muon-electron mass ratio 
does involve experiments with \mm\ \cite{liu99}, 
but it is performed using spin-dependent transitions 
that can be considered independent
of the isotropic nonrelativistic coefficients of interest here.

Taking the difference between the experimental
and theoretical values gives 
$\nu_{1S2S}^{\rm expt} -\nu_{1S2S}^{\rm th} = 5.3\pm9.8$ MHz.
We interpret this conservatively as implying the 
difference is zero to within $\pm$20 MHz,
yielding the bound on a combination 
of isotropic nonrelativistic coefficients 
given by 
\bea
\big|\anrfc{2}+ \cnrfc{2} 
+(8\times 10^{-11} {\rm ~GeV}^2)
( \anrfc{4}+ \cnrfc{4} ) \big|
&&
\nonumber\\
&&
\hskip -80pt
<8\times 10^{-6}{\rm ~GeV}^{-1}.
\qquad
\label{1s2sIso}
\eea
Note that isotropy implies this bound holds unchanged 
when the coefficients are evaluated in the Sun-centered frame. 
The resulting constraints on individual coefficients
taken one at a time
are listed in the first four rows of Table \ref{table3}.

Another interesting option for studying 
the effects of isotropic nonrelativistic coefficients
is the splitting $2S_{1/2}$-$2P_{1/2}$,
which is the Lamb shift in \mm.
The shift $\de\nu_{\rm Lamb}$
of the Lamb frequency $\nu_{\rm Lamb}$
due to Lorentz and CPT violation
can be obtained by noting that the perturbative hamiltonian \rf{1S} 
applies for both the $2S_{1/2}$ and the $2P_{1/2}$ levels
and by using Eq.\ \rf{radialExp}
to calculate the appropriate expectation values.
Restricting attention to the spin-independent terms,
we thereby obtain
\beq
2\pi\de \nu_{\rm Lamb}=
-\frac{2}{3}(\al \mr)^4
\left( \anrfc{4}+ \cnrfc{4}\right).
\label{ML}
\eeq
As before,
we can estimate a bound on this effect
by comparing
the experimental and theoretical values,
$\nu_{\rm Lamb}^{\rm expt} = 1042^{+21}_{-23}$ MHz 
and
$\nu_{\rm Lamb}^{\rm th} = 1047.490(300)$ MHz
\cite{woodle}.
Taking the Lorentz-violating effect as smaller than $\pm$30 MHz
gives the conservative bound
\beq
|\anrfc{4}+ \cnrfc{4}| < 1\times 10^6 {\rm ~GeV}^{-3},
\label{Lambm}
\eeq
which also holds in the Sun-centered frame.
The resulting constraints
on each of the two isotropic nonrelativistic coefficients taken in turn
are given in the fifth and sixth rows of Table \ref{table3}. 

\renewcommand\arraystretch{1.5}
\begin{table}
\caption{Constraints on muon isotropic nonrelativistic coefficients
from \mm\ spectroscopy.}
\setlength{\tabcolsep}{5pt}
\begin{tabular}{ccl}
\hline
\hline
Transition & Coefficient & Constraint\\
\hline
$1S_{1/2}$-$2S_{1/2}$ &
$|\anrfc{2}|$
& $< 8\times 10^{-6}$ GeV$^{-1}$\\
&$|\cnrfc{2}|$
& $< 8\times 10^{-6}$ GeV$^{-1}$\\
&$|\anrfc{4}|$
& $<1 \times 10^5$ GeV$^{-3}$\\
&$|\cnrfc{4}|$ 
& $< 1 \times 10^5$ GeV$^{-3}$\\
Lamb shift &
$|\anrfc{4}|$
& $< 1\times 10^6$ GeV$^{-3}$\\
&$|\cnrfc{4}|$ 
& $< 1\times 10^6$ GeV$^{-3}$\\
\hline
\hline
\end{tabular}
\label{table3}
\end{table}

\subsection{Muonic atoms and ions}
\label{Muonic hydrogen}

Next,
we investigate the use of spectroscopy of muonic atoms and ions 
to study Lorentz and CPT violation.
The focus in the first few subsections that follow
is primarily on the \hm\ transitions 
$2S^{F-1}_{1/2}$-$2P^F_{3/2}$ with $F=1,2$.
These have recently been measured 
at the Paul Scherrer Institute (PSI)
\cite{antognini},
leading to the proton radius puzzle
\cite{pohl}.
In Sec.\ \ref{Other muonic atoms},
the analogous transitions in various other muonic atoms and ions
are discussed.
In what follows,
we identify signals for Lorentz violation 
that can be sought in experiments for all these systems,
and we investigate the prospects 
for resolving the proton radius puzzle
via Lorentz and CPT violation.

\subsubsection{Generalities}
\label{Generalities}

To enable comparisons between \hm\ and \mm\ experiments,
it is convenient to consider
scenarios with similar relative precision
and use scaling arguments to determine the relevant absolute precision.
Thus,
for example,
if the relative precisions of the $1S$-$2S$ frequency 
in \hm\ and \mm\ are roughly comparable,
then the energies and hence the absolute precisions 
are scaled by the ratio $\simeq$187 of the reduced masses. 
For instance,
a 2 GHz sensitivity in \hm\
corresponds to roughly the same relative precision
as a 10 MHz sensitivity in \mm. 
This energy scaling affects the reach
of searches for Lorentz and CPT violation.
The experimental sensitivity to coefficients with $d=3$ 
is determined by the absolute frequency resolution
and is therefore reduced in \hm\ experiments,
though it can still be of definite interest
in the absence of other available results.
Also,
\hm\ and \mm\ experiments with similar relative precision
should have comparable sensitivity to dimensionless coefficients. 
However,
for $d\geq 5$
the reach of \hm\ experiments can be expected 
to be superior by a factor of about $(187)^{d-4}$.

Many of the derivations for \mm\ 
in the previous subsection
can be adapted for \hm,
and the attainable sensitivities can crudely be estimated 
using similar reasoning.
For example,
studies of sidereal variations of the $1S$ hyperfine splittings in \hm\
would be of definite interest
and can be expected to lead to sensitivities to 
$\HzBnr{211}$, $\HoBnr{211}$, $\gzBnr{211}$, $\goBnr{211}$
improved by about an order of magnitude
and sensitivities to
$\HzBnr{411}$, $\HoBnr{411}$, $\gzBnr{411}$, $\goBnr{411}$
improved by more than five orders of magnitude.

One primary interest is the Lamb shift in \hm,
and in particular the transitions
$2S^{F-1}_{1/2}$-$2P^F_{3/2}$
with $F=1,2$.
Since $F\le2$, 
only SME coefficients with $j\le 3$ 
can contribute to spin-dependent effects.  
However,
sensitivity to all such effects is better by many orders of magnitude
in experiments studying the muon anomalous magnetic moment,
as is verified in Sec.\ \ref{Muon magnetic moment} below.
The essential point is that in both types of experiments 
the unperturbed system has even parity
while the relevant spin-dependent Lorentz-violating operators 
have parity $(-1)^{j+1}$,
so only coefficients with odd $j$ contribute.
We can therefore reasonably disregard spin-dependent effects
in the present context. 

For spin-independent effects,
only the coefficients with $j\le 2$ can contribute 
because the total angular momentum of the muon 
is $J\le 3/2$.
Neglecting any Lorentz violation in the proton sector,
which in any event can be better studied using conventional atoms,
the perturbative terms relevant for the Lamb shift
can therefore be taken as 
\bea
\de h^\nr_{2S2P}&=&
\sum_{q=0}^2\pmag^{2q} \Bigg[ 
(\anrfc{2q}-\cnrfc{2q})
\nonumber\\ 
&& 
\hskip 30pt
+\sum_{m=-2}^2\left(\anr{(2q)2m}-\cnr{(2q)2m}\right)
Y_{2m}(\phat)\Bigg].
\nonumber\\ 
\label{2P}
\eea
The coefficients $\anrfc{2q}$ and $\cnrfc{2q}$ have $j=0$ 
and so affect both the $2S^F_{1/2}$ and the $2P^F_{3/2}$ states
isotropically.
The other terms have $j=2$ 
and contribute only to the $2P_{3/2}^{F}$ levels,
with the shift varying with the orientation 
of the total angular momentum $\mbf F$.

\subsubsection{Zeeman transitions}
\label{magnetic field}

If the Zeeman splittings due to the applied magnetic field 
are larger than those due to Lorentz violation,
then sidereal variations of the Lamb transition frequencies occur.
Searching for these effects in \hm\ requires resolving the Zeeman shift
and accumulating sufficient statistics to perform sidereal studies,
which is unrealized to date but may be possible in future experiments.
 
The frequency shift $\de\nu (F,m_F)$
of the $2S^{F-1}_{1/2}$-$2P^F_{3/2}$ transition at fixed $m_F$
induced by Lorentz violation 
involves the expectation value
\beq
\vev{F,m_F|Y_{20}(\phat)|F,m_F}=
\fr{(5-2F)(F(F+1)-3m_F^2)}{12\sqrt{5\pi}},
\label{E3/20}
\eeq
where the state $\ket{F,m_F}$ is understood to have
$J=3/2$ and $L=1$.
Using this result and the perturbation hamiltonian \rf{2P},
we obtain 
\bea
2\pi \de \nu(F,m_F)&=& 
\frac 23 (\al m_{\rm r})^4
(\cnrfc{4}-\anrfc{4}) 
\nonumber\\ 
&& 
+ \fr{(5-2F)(F(F+1)-3m_F^2)}{12\sqrt{5\pi}}
\q_{20},
\qquad
\label{wm1}
\eea
where for convenience we define 
\beq
\q_{jm}\equiv\sum_{k=0}^{2}\vev{\pmag^{2k}}_{21}
(\anr{(2k)jm}-\cnr{(2k)jm}).
\eeq

To display explicitly its sidereal time dependence,
the frequency shift $\de \nu(F,m_F)$ can be expressed
in terms of coefficients in the Sun-centered frame.
Using the Wigner rotation matrices 
\cite{wigner}
$D^{(j)}_{m m^\prime}(\al,\be,\ga)$
to transform between frames
shows that
the laboratory-frame combination $\q_{20}^{\rm lab}$ 
is related to coefficients $\q_{jm}^{\rm Sun}$ 
in the Sun-centered frame by
\cite{km09}
\bea
\q_{20}^{\rm lab}&=&
\sum_{m=-2}^2  D_{0m}^{(2)}(0,-\ch,-\om_\oplus T_\oplus)\q_{2m}^{\rm Sun}
\nonumber \\ 
&=& \sqrt{\fr{4\pi}{5}}
\sum_{m=-2}^2  Y_{2m}(\ch,\om_\oplus T_\oplus) \q_{2m}^{\rm Sun}
\label{sd2}
\eea
where 
$\ch$ is the angle between the magnetic field 
and the rotational north pole of the Earth,
$\om_\oplus$ is the Earth sidereal frequency,
and $T_\oplus$ is the sidereal time,
as before. 
This expression implies that in the Sun-centered frame
the frequency shift $\de \nu(F,m_F)$ 
takes the form 
\bea
2\pi \de\nu(F,m_F)&=& 
\frac 23 (\al m_{\rm r})^4
(\cnrfc{4}-\anrfc{4}) 
\nonumber\\ 
&& 
+ \fr{(5-2F)(F(F+1)-3m_F^2)}{30} 
\nonumber\\ 
&& 
\hskip 20pt
\times \sum_{m=-2}^2 Y_{2m}(\ch,\om_\oplus T_\oplus) \q_{2m},
\qquad
\label{wm2}
\eea
where all coefficients are now expressed in the Sun-centered frame.  

The result \rf{wm2} reveals that 
future experiments sensitive to the Zeeman shift in Lamb transitions 
can be used to search for Lorentz violation 
through sidereal variations.
The coefficients
$\anr{221}$, $\anr{421}$, $\cnr{221}$, $\cnr{421}$ 
control oscillations at the sidereal frequency $\om_\oplus$,
while
$\anr{222}$, $\cnr{222}$, $\anr{422}$, $\cnr{422}$ 
control ones at $2\om_\oplus$.
Suppose,
for example,
measurements are made for $F=1$, $m_F=0$,
corresponding to an experiment  
with a laser polarized in the direction of the magnetic field.
Assume the magnetic field is inclined at $\ch=45^\circ$ 
to the Earth's rotation axis
and the experiment establishes no sidereal signal at $\pm1$ GHz.
Then,
constraints of order $10^{-7}$ GeV$^{-1}$ 
could be placed on $|\anr{22m}|$, $|\cnr{22m}|$
and ones of order 1 GeV$^{-3}$ 
on $|\anr{42m}|$, $|\cnr{42m}|$.
A comparable \mm\ counterpart experiment
would achieve a resolution of about $\pm5$ MHz,
with corresponding sensitivities some two orders of magnitude weaker 
on $|\anr{22m}|$, $|\cnr{22m}|$
and about seven orders of magnitude weaker
on $|\anr{42m}|$, $|\cnr{42m}|$.

\subsubsection{Proton radius puzzle}
\label{Proton radius puzzle}

The result \rf{wm2} for the shifts 
in the $2S^{F-1}_{1/2}$-$2P^F_{3/2}$ transition frequencies 
contains terms with $m=0$
that are independent of sidereal time.
These constant shifts 
can be expected to appear 
as a discrepancy between experimental measurements
and conventional Lorentz-invariant theoretical predictions. 
However,
the theoretical predictions depend 
on the value of the proton charge radius $r_p$,
so in practice
the recent PSI measurements of these transitions
are used to extract an independent measure of $r_p$ instead 
\cite{antognini}.
Surprisingly,
this measure is in apparent disagreement by about seven standard deviations
with the 2010 CODATA value
obtained by combining results from hydrogen spectroscopy
and from electron elastic scattering data
\cite{mohr}.
The difference $\De r_p \simeq -0.037\pm 0.005$ fm between these values
suggests a smaller proton radius measured by \hm\ spectroscopy,
a result called the proton radius puzzle
\cite{pohl}.

Since Lorentz violation can induce a constant shift
in the Lamb transition frequencies
and hence an apparent constant shift
in the inferred proton charge radius,
we can ask what size Lorentz violation 
would suffice to resolve the puzzle
and what implications this might have for other experiments.
For simplicity we assume only muon-sector Lorentz violation as before,
so effects arise in \hm\ spectroscopy
but are absent in H spectroscopy and electron elastic scattering. 
We also disregard any effects from the sidereal variations
discussed in the previous subsection.
A more complete analysis could conceivably
demonstrate a sidereal-time dependence
in the inferred value of the proton charge radius. 

Taking for definiteness the polarization of the laser 
in the direction of the magnetic field,
for which $\Delta m_F=0$, 
and assuming an equal population 
of the initial states with $m_F= -1$, 0, 1, 
we find the induced change in the Lamb-shift energy is given by 
\beq
\de E_{\rm Lamb} = 
\frac 23 (\al m_{\rm r})^4
(\cnrfc{4}-\anrfc{4}) 
+ \fr{3(1+3\cos{2\ch)}}{32\sqrt{5\pi}}\q_{20}, 
\eeq
with 
\beq
\q_{20}=
\frac 14 (\al m_r)^2 (\anr{220}-\cnr{220})
+ \frac 7{48} (\al m_r)^4
(\anr{420}-\cnr{420}).
\eeq
All SME coefficients appearing in these equations have $m=0$
and are expressed in the Sun-centered frame.

The theory relating the Lamb shift to the proton charge radius $r_p$
\cite{antognini,eides,pcrth,antognini2}
implies that $\de E_{\rm Lamb}$ 
can be interpreted as a change $\de r_p$ 
in the determination of $r_p$ given by
\beq
\de r_p ~({\rm fm})
\simeq
- 1.1\times 10^{11}
\de E_{\rm Lamb}~({\rm GeV}).
\eeq
Within the present hypothesis
attributing the discrepancy $\De r_p$ 
to the shift $\de r_p$ induced by Lorentz violation,
we can impose $\De r_p = \de r_p$
and thereby establish the requirement 
for the coefficients for Lorentz violation
to resolve the proton radius puzzle. 
This gives the condition
\bea
\frac 23 (\al m_{\rm r})^4
(\cnrfc{4}-\anrfc{4}) 
+ \fr{3(1+3\cos{2\ch)}}{32\sqrt{5\pi}}\q_{20} 
\nonumber\\
&&
\hskip -100 pt
\simeq 3\times 10^{-13} {\rm ~GeV}.
\qquad
\label{Br}
\eea
Any combination of coefficients satisfying this equation
would suffice to resolve the discrepancy in the charge radius.
Note that these results depend on the chosen angle $\ch$ 
between the magnetic field and the Earth's rotation axis.
Using these coefficients to resolve the proton radius puzzle
therefore comes with a prediction
that the inferred value of $r_p$ 
could vary with the orientation of the magnetic field.

The PSI experiment also 
deduces the $2S$ hyperfine splitting
and hence determines the Zemach magnetic radius $r_Z$ of the proton
\cite{antognini}.
The result is in agreement with data from H spectroscopy
and from electron-proton scattering,
and the difference $\De r_Z$ between them 
can conservatively be taken as bounded by $|\De r_Z| < 0.07$ fm.
This result places an additional constraint
on the coefficients appearing in Eq.\ \rf{Br}
because some of them also affect 
the determination of the hyperfine splitting.
In the same scenario as before,
we find the Lorentz-violating shift $\de E_{\rm HF}$
in the hyperfine interval to be
\beq
\de E_{\rm HF} = 
\fr{(1+3\cos{2\ch)}}{24\sqrt{5\pi}}\q_{20}. 
\eeq
The theory relating the hyperfine splitting to the Zemach radius
\cite{antognini,pzrth,antognini2}
shows that the shift $\de E_{\rm HF}$  
can be understood as a change $\de r_Z$
given as
\beq
\de r_Z ~({\rm fm}) \simeq 
- 6.2 \times 10^{12} 
\de E_{\rm HF} ~({\rm GeV}).
\eeq
Following analogous reasoning as before,
we can impose $\De r_Z = \de r_Z$
to obtain the constraint on SME coefficients
required to preserve the agreement between the various experiments
determining the Zemach radius.
This gives
\beq
\Big|
\fr{(1+3\cos{2\ch)}}{24\sqrt{5\pi}}\q_{20}
\Big|
\lsim 1 \times 10^{-14} {\rm ~GeV}.
\label{zb}
\eeq
This condition is tighter than the constraint \rf{Br}
by about an order of magnitude.
It depends on the orientation $\ch$ of the magnetic field
in the experiment,
which may be worth exploring experimentally.
Note that the same linear combination of coefficients $\q_{20}$ 
appears in both conditions \rf{Br} and \rf{zb}.
Within the present scenario,
this means that resolving the proton radius puzzle
via Eq.\ \rf{Br}
requires a nonzero value for the combination 
$\cnrfc{4}-\anrfc{4}$
of isotropic coefficients,
a possibility remaining compatible with existing constraints.

Some additional intuition for the implications of these results
can be gained by extracting from Eq.\ \rf{Br}
the corresponding condition on each coefficient in turn
with all others set to zero.
This procedure gives 
$\cnrfc{4} \simeq 2$ GeV$^{-3}$
and $\anrfc{4} \simeq - 2$ GeV$^{-3}$
for the isotropic coefficients taken one at a time.
For the coefficients with $j=2$ and taking $\ch = 45^\circ$,
we find
$\anr{220} = - \cnr{220} = 10^{-4}$ GeV$^{-1}$
and 
$\anr{420} = - \cnr{420} = 400$ GeV$^{-3}$.
Comparison with the results in Table \ref{table3}
reveals that $1S$-$2S$ and 
Lamb-shift spectroscopy in \hm\ 
is several orders of magnitude more sensitive
to the isotropic coefficients than the corresponding measurement in \mm.
We remark in passing that
the distinction between using $\anrfc{4}$ and $\cnrfc{4}$
in this context is experimentally undetectable
as it would require spectroscopic studies of \hmb,
for which the coefficient $\anrfc{4}$ for CPT-odd effects
would cause an apparent increase of the antiproton charge radius.

\subsubsection{Negligible magnetic field}
\label{Negligible magnetic field}

For present purposes,
the $\simeq 5$ T magnetic field used
in the recent PSI experiment
\cite{antognini}
can be viewed as negligible
because the Zeeman splitting remains unobserved.
Disregarding for the moment the laser polarization,
the apparatus itself can be idealized as rotationally invariant
in the laboratory frame.
This implies that observable sidereal variations cannot appear.
However,
the presence of Lorentz violation
acts to break the rotational symmetry of the \hm\ atom
and partially or wholly lifts the $(2F+1)$-fold degeneracy 
with respect to the orientation of its total angular momentum $\mbf F$.
This offers an alternative route to searching for Lorentz violation,
as we describe next.

The energy splittings resulting from Lorentz violation
depend on the value of $F$,
and the corresponding $2F+1$ states can be labeled
by an effective azimuthal quantum number $\xm$
taking the values $-F$, $-F+1$, $\ldots$, $F-1$, $F$, 
as usual.
Note that we use $\xm$ instead of $m_F$ here
to avoid possible confusion 
with projection of $\mbf F$ along the magnetic field. 
The energy shift $\de E(F,\xm)$ for a given state 
can be obtained from the perturbation \rf{2P},
which depends on coefficients for Lorentz violation
having either $j=0$ or $j=2$. 
The coefficients with $j=0$ govern isotropic Lorentz violation,
so only those with $j=2$ can shift the levels according to 
the orientation of $\mbf F$.
As discussed above,
only states with $F=1$ or $F=2$ are affected,
and in particular for the $2S^{F-1}_{1/2}$-$2P^F_{3/2}$ transitions
only the $P$ states are relevant.
We therefore focus in what follows
on the shift of the $P$ states
controlled by coefficients with $j=2$.

For an arbitrary orientation of $\mbf F$
in the laboratory frame,
the explicit form of the level shift $\de E(F,\xm)$ 
is involved,
being determined by the solution of a cubic for $F=1$
and by a quintic for $F=2$.
For example,
for $F=1$ we obtain
\beq
\de E(1, \xm)=
\frac 23 (\al m_{\rm r})^4
(\cnrfc{4}-\anrfc{4}) 
+ u_\xm D
+u^*_\xm \dfrac{\De_0}{D},
\label{mHF1}
\eeq
where 
\beq
u_\xm=\fr{{1+ i \xm \sqrt{3}}}{3(1-3\xm^2)},
\qquad
\eeq
and 
\beq
D= \left(\dfrac{\De_1+\sqrt{\De_1^2-4\De_0^3}}{2}\right)^{1/3}.
\eeq
The quantities $\De_0$ and $\De_1$  
contain the combinations of coefficients for Lorentz violation
defined in Eq.\ \rf{sd2}
and are given by 
\bea
\De_0&=&
\fr 9{80\pi}\sum_{m=-2}^2 |\q_{2m}|^2,
\nonumber\\
\De_1&=& 
-\fr{27\q_{20}}{160\sqrt{5\pi^3}}
\left(6|\q_{22}|^2-(\q_{20})^2-3|\q_{21}|^2\right)
\nonumber\\
&&
+\fr{81}{80}\sqrt{\fr{3}{10\pi^3}}
\Re\left[\q_{22}^*(\q_{21})^2\right]. 
\label{D1}
\eea
We note in passing that all the quantities appearing in these equations
are rotational scalars,
so the above expressions 
hold at leading order both in the laboratory frame
and in the Sun-centered frame.

Some insight into the content of the results for both $F=1$ and $F=2$
can be obtained by extracting the level shifts $\de E(F,\xm)$ 
under the assumption that only one combination of coefficients $\q_{jm}$ 
is nonzero at a time.
Table \ref{Spec2} displays the spectral shifts
obtained in this scenario.
Inspection of the table reveals that the coefficient $\q_{20}$ 
lifts the degeneracy only partially,
while $\q_{2m}$ with $m\neq 0$ lifts it completely.  

\renewcommand\arraystretch{2.3}
\begin{table}
\caption{Spectral shifts $\de E(F,\xm)$ for $j=2$.}
\setlength{\tabcolsep}{8pt}
\begin{tabular}{cccc}
\hline\hline
&&&\\
[-30pt]
$F$ & $\xm$ & $m=0$ & $m\neq 0$ \\
\hline
1 & 1 &
$-\dfrac{\q_{20}}{4\sqrt{5\pi}}$
&
$\sqrt{\dfrac{3}{10\pi}}
\dfrac{|\q_{2m}|}{2}$
\\
& 0 &
$\dfrac{\q_{20}}{2\sqrt{5\pi}}$
&
0
\\
& $-1$ &
$-\dfrac{\q_{20}}{4\sqrt{5\pi}}$
&
$-\sqrt{\dfrac{3}{10\pi}} \dfrac{|\q_{2m}|}{2}$
\\
2 & 2 &
$-\dfrac{\q_{20}}{2\sqrt{5\pi}}$
&
$\dfrac{|\q_{2m}|}{\sqrt{10\pi}}$
\\
& 1 &
$\dfrac{\q_{20}}{4\sqrt{5\pi}}$
&
$\sqrt{\dfrac{3}{10\pi}} \dfrac{|\q_{2m}|}{2}$ 
\\
& 0 &
$\dfrac{\q_{20}}{2\sqrt{5\pi}}$
&
0
\\
& $-1$ &
$\dfrac{\q_{20}}{4\sqrt{5\pi}}$
&
$-\sqrt{\dfrac{3}{10\pi}} \dfrac{|\q_{2m}|}{2}$ 
\\
& $-2$ &
$-\dfrac{\q_{20}}{2\sqrt{5\pi}}$
&
$-\dfrac{|\q_{2m}|}{\sqrt{10\pi}}$
\\
[4pt]
\hline\hline
\end{tabular}
\label{Spec2}
\end{table}

To illustrate how the spectral splitting
can be used to seek Lorentz violations,
suppose that all transitions 
$2S_{1/2}^{F=0}$-$2P_{3/2}^{F=1,\xm}$ 
are excited with equal probability,
perhaps via an unpolarized laser.
An experimental measurement 
effectively involves a cloud of atoms with random 
orientations $\mbf F$,
so the Lorentz violation acts to broaden 
the observed transition line.
The theoretical apparent width $\De E$ of the $2P$ level
due to Lorentz violation 
is given by
\beq
(\De E)^2 =
\frac 13\sum_{\xm=-1}^1 \de E(1,\xm)^2
-\frac 19 \left(\sum_{\xm=-1}^1 \de E(1,\xm)\right)^2,
\eeq
from which the result 
\beq
(\De E)^2=\frac 29\De_0
=\fr{1}{40\pi}\sum_{m=-2}^2 |\q_{2m}|^2
\label{combo}
\eeq
is obtained.

Assuming the experiment cannot directly resolve the splitting,
it follows that the precision of the measurement 
can be taken as an upper bound on $\De E$,
thereby yielding upper limits on the combination of coefficients
given in Eq.\ \rf{combo}. 
This combination contains 
12 independent complex nonrelativistic coefficients.
As before,
a measure of the attainable sensitivity
can be obtained by taking each coefficient nonzero in turn. 
For example,
an experiment achieving a precision of about 1 GHz
would place constraints on $|\anr{22m}|$, $|\cnr{22m}|$
at the level of about $10^{-7}$ GeV$^{-1}$,
while the constraints on on $|\anr{42m}|$, $|\cnr{42m}|$
would be at the level of about 1 GeV$^{-3}$.
A \mm\ experiment with comparable fractional sensitivity 
would have a 5 MHz precision,
but it would yield constraints
some two to seven orders of magnitude weaker on the same coefficients.

In realistic applications,
the laser polarization is fixed in the laboratory frame.
For example,
in the PSI experiment
\cite{antognini},
the polarization is parallel to the magnetic field. 
The plane of the polarization rotates with the Earth,
inducing sidereal oscillations in the presence of Lorentz violation.
A detailed analysis would therefore involve
a combination of the line-broadening effects described above 
with sidereal oscillations associated with the laser polarization. 

\subsubsection{Other muonic atoms and ions}
\label{Other muonic atoms}

A variety of other muonic atoms and ions
can be used to search for signals of Lorentz and CPT violation.
The PSI experiment R98-03 has  
already measured several transitions in \dm\ 
\cite{deuterium},
while the 
the Charge Radius Experiment with Muonic Atoms (CREMA),
PSI project R10-01,
proposes to study the Lamb shift 
in \hetm\ and \hefm\ ions
\cite{crema}.
Possibilities may also exist for muonic tritium \tm\
\cite{fujiwara}
and for the heavier ions 
\lixm, \linm, \bem, and \bom\
\cite{drakebyer}.
The key differences in the hydrogenic spectra
of all these muonic systems arise through differences 
in the nuclear spin $I$, the net charge $Z$, 
and the reduced mass $m_{\rm r}$.
This makes possible a unified treatment
of the Lorentz-violating corrections to their Lamb shifts,
following the methods established for \hm\
in the previous subsections.

Consider first for any one of these systems
the analogue of the Zeeman transitions
described in Sec.\ \ref{magnetic field},
which may become accessible to future experiments.
For example,
in \dm\ the nucleus has spin $I=1$,
so the transitions of interest are 
$2S^{F-1}_{1/2}$-$2P^{F}_{3/2}$ at fixed $m_F$
with $F\in\{3/2,5/2\}$.
Similarly,
for \hetm\ the spin is $I=1/2$,
so the transitions are like those of \hm. 
For \hefm\ the spin is $I=0$,
so $F=J$ and the transitions of interest are 
$2S_{1/2}$-$2P_{1/2}$ and $2S_{1/2}$-$2P_{3/2}$ 
at fixed $m_F$.

The Lorentz-violating shift in the frequency
for any of these systems is given by a generalization
of Eq.\ \rf{wm1},
\bea
2\pi \de \nu(F,m_F)&=& 
\frac 23 (Z\al m_{\rm r})^4
(\cnrfc{4}-\anrfc{4}) 
\nonumber\\ 
&& 
+ \frac{1}{\sqrt{5\pi}}
\La(F)(F(F+1)-3m_F^2) \q_{20}^\prime,
\qquad
\label{genres}
\eea
where
$\La(F)$ is a factor specific to the atom or ion
and where 
\beq
\q_{20}^\prime =
\frac 14(Z\al m_{\rm r})^2 (\anr{220}-\cnr{220})
+\frac{7}{48} (Z\al m_{\rm r})^4 (\anr{420}-\cnr{420}).
\eeq
The values taken by $\La(F)$ are displayed in Table \ref{LF}.

To gain some insight into the implications of Eq.\ \rf{genres},
it is useful first to establish a relative measure 
of spectroscopic precision for the various systems.
Since the Lamb shift is proportional to $Z^4 m_{\rm r}$, 
comparing different experiments 
assuming the same relative uncertainty 
implies comparing ratios of this factor. 
For example,
the ratio of the factors $Z^4 m_{\rm r}$ 
for \hefm\ and \hm\ is about 20, 
so a precision of 1 GHz in \hm\ 
is comparable to a precision of 20 GHz in \hefm.  

\renewcommand\arraystretch{1.5}
\begin{table}
\caption{Lamb-shift factors $\La(F)$ and 
relative sensitivities of muonic atoms and ions
to Lorentz and CPT violation.
The estimated relative attainable sensitivities 
to the SME coefficients $\anr{kj0}$ and $\cnr{kj0}$ 
are shown for different values of $kj0$, 
using a normalization relative to \hm.}
\setlength{\tabcolsep}{2pt}
\begin{tabular}{cccccccc}
\hline\hline
&	&	&	&	&	&	$kj0$	&	\\
[-4pt]
System	&	$I$	&	$Z$	&$m_{\rm r}$ (GeV)	&	$\La(F)$		&	400	&	220	&	420	\\
\hline
\hm	&	$\half$	&	1	&	0.094	&	$\frac{1}{12}(5-2F)$	&	1	&	1	&	1	\\
\dm	&	1	&	1	&	0.099	&	$\frac{1}{120}(2F+1)$	&	0.85	&	4.7	&	4.3	\\
\tm	&	$\half$	&	1	&	0.101	&	$\frac{1}{12}(5-2F)$	&	0.81	&	0.93	&	0.81	\\
\hetm	&	$\half$	&	2	&	0.101	&	$\frac{1}{12}(5-2F)$	&	0.81	&	3.7	&	0.81	\\
\hefm	&	0	&	2	&	0.102	&	$\frac{1}{12}(2F-1)$	&	0.79	&	3.7	&	0.79	\\
\lixm	&	1	&	3	&	0.103	&	$\frac{1}{120}(2F+1)$	&	0.77	&	41	&	3.8	\\
\linm	&	$\frac32$	&	3	&	0.103	&	$\frac{1}{60}(F-2)(17-5F)$\quad	&	0.77	&	10	&	0.95	\\
\bem	&	$\frac32$	&	4	&	0.104	&	$\frac{1}{60}(F-2)(17-5F)$\quad	&	0.75	&	18	&	0.94	\\
\bom	&	$\frac32$	&	5	&	0.104	&	$\frac{1}{60}(F-2)(17-5F)$\quad	&	0.75	&	28	&	0.94	\\
[5pt]
\hline\hline
\label{LF}
\end{tabular}
\end{table}

With this measure in hand,
we can provide estimates of the relative sensitivities
of each muonic system
to the coefficients for Lorentz violation
appearing in Eq.\ \rf{genres}.
For the coefficients with $k=2$,
the relevant factor is $Z^2/m_{\rm r}$,
while for coefficients with $k=4$
it is $1/m_{\rm r}^3$.
Table \ref{LF} shows the resulting estimated sensitivities
of the different muonic systems
relative to \hm\
for the SME coefficients $\anr{kj0}$ and $\cnr{kj0}$
with $kj0=$400, 220, and 420.
The reader is reminded that
$\anr{k00} \equiv \sqrt{4\pi} \anrfc{k}$ and
$\cnr{k00} \equiv \sqrt{4\pi} \cnrfc{k}$.
The entries in this table assume the smallest possible value
of $F$ allowed in each case.
A numerical value less than one implies 
that Lamb-shift spectroscopy using the corresponding system
is estimated to be more sensitive by that value than 
spectroscopy using \hm. 

Table \ref{LF} demonstrates that 
future experiments studying the Lamb-shift Zeeman transitions 
in various muonic atoms and ions
can achieve interesting sensitivities
to nonrelativistic coefficients.
In the Sun-centered frame,
the transition frequencies acquire a sidereal time dependence
following Eq.\ \rf{sd2},
so experiments can measure
the coefficients $\anr{221}$, $\anr{421}$, $\cnr{221}$, $\cnr{421}$ 
via the sidereal frequency $\om_\oplus$
and the coefficients $\anr{222}$, $\cnr{222}$, $\anr{422}$, $\cnr{422}$ 
via $2\om_\oplus$.
The sensitivies relative to those of \hm\ 
can be obtained from the table.
Assuming,
for example,
that an experiment with \dm\
detects no sidereal signal at the level of 2 GHz
for the transitions with $F=3/2$, $m_F=1/2$
using a laser polarized along 
the magnetic field orientation $\ch=45^\circ$,  
then constraints of order $10^{-6}$ GeV$^{-1}$ 
could be placed on $|\anr{22m}|$, $|\cnr{22m}|$
and ones of order 1 GeV$^{-3}$ 
on $|\anr{42m}|$, $|\cnr{42m}|$.
A similar experiment with \hefm\
with no sidereal signal at 20 GHz
would achieve roughly comparable sensitivities on coefficients with $k=2$
and about a factor of 5 improvement on coefficients with $k=4$,
with the latter gain being primarily 
due to the larger reduced mass.

We can also consider the case of negligible magnetic field
discussed in Sec.\ \ref{Negligible magnetic field},
where the Zeeman splittings are unresolved.
Lorentz and CPT violation in any of the systems
in Table \ref{LF} then again appears
as a line broadening resulting from the breaking of rotational symmetry
and the associated dependence on the orientation of 
the total angular momentum $\mbf F$.
The apparent width $\De E$ can be found
in each case using the techniques leading to Eq.\ \rf{combo}.

Limiting attention to the isotropic muon coefficients,
the Lorentz-violating shift in the Lamb energy 
for any of the muonic atoms or ions is given by 
\beq
\de E_{\rm Lamb} = 
\frac 23 (Z\al m_{\rm r})^4
(\cnrfc{4}-\anrfc{4}) .
\eeq
Table \ref{LF} shows that the heavier systems 
are more sensitive to isotropic coefficients than \hm.
Note that the net effect in each case would be an apparent shift 
in the nuclear charge radius of the muonic atom or ion
relative to the equivalent electron system.
In particular,
it might be possible with these experiments
to exclude or confirm
any contribution to the proton radius puzzle 
arising from isotropic coefficients
as discussed in Sec.\ \ref{Proton radius puzzle}.

Finally,
we note that another signal of Lorentz violation 
is a change in the isotope shifts
between \hm\ and the other muonic atoms or ions
relative to the isotope shifts of the corresponding electron systems.
Again considering only the isotropic muon coefficients,
the 1S-2S transition frequency in a given atom or ion is shifted by 
\beq
2\pi \de \nu = 
\frac 34 (Z\al m_{\rm r})^2 (\cnrfc{2}-\anrfc{2}) 
+ \frac {67}{16} (Z\al m_{\rm r})^4 (\cnrfc{4}-\anrfc{4}) .
\eeq
This result implies an apparent isotope shift 
arising from the nuclear charge and the reduced mass,
and appearing only in experiments with muonic systems.

\section{Muon magnetic moment}
\label{Muon magnetic moment}

In this section,
the effects of Lorentz and CPT violation
on measurements of the muon anomalous magnetic moment are considered.
The muon anomaly frequency,
which in Lorentz-invariant models is proportional to 
the muon $g-2$ factor,
was studied over a 20-year period
in a series of experiments at CERN
\cite{cern}.
More recently,
it has been measured 
to an impressive precision of about 0.5 ppm
in experiment E821 at the Brookhaven National Laboratory (BNL) 
\cite{bnl}.
The upcoming experiment E989 
at the Fermi National Accelerator Laboratory (Fermilab)
\cite{fermilab} 
and an experiment at the Japan Proton Accelerator Research Complex (J-PARC) 
\cite{j-parc} 
both anticipate roughly a fivefold improvement over this mark. 
In the presence of Lorentz violation,
the muon and antimuon anomaly frequencies $\om_a^-$ and $\om_a^+$ 
can acquire a difference,
and sidereal and annual variations of $\om_a^\pm$ can also appear.
In what follows,
we outline the underlying basis for these effects 
and then consider each of the resulting types of signals in turn.

\subsection{Basics}
\label{Theory}

In the BNL experiment
\cite{bnl},
relativistic polarized $\mu^+$ or $\mu^-$ beams 
were injected into cyclotron orbits 
in a constant magnetic field $B\simeq$ 1.45 T 
and adjusted to the `magic' momentum $p \simeq$ 3.094 GeV 
with $\ga\simeq 29.3$ 
at which the dependence of the anomaly frequency
on the electric field is eliminated. 
Fitting the $\mu^\pm$ decay spectrum
permits inferring the corresponding anomaly frequency $\om_a^\pm$,
which is the difference between
the spin-precession frequency $\om_s^\pm$ 
and the cyclotron frequency $\om_c^\pm$.
The earlier CERN experiment
\cite{cern}
and the upcoming Fermilab experiment
\cite{fermilab}
involve designs conceptually similar to the BNL one.
In contrast,
the J-PARC experiment
\cite{j-parc}
will use ultracold highly polarized $\mu^+$ beams 
of momentum $p\simeq 320$ MeV and $\ga\simeq 3.03$ 
that can be stored in a magnetic field $B\simeq$ 3 T
without a focusing electric field. 

In all these experimental scenarios,
the leading-order corrections to the anomaly frequencies $\om_a^\pm$
arising from Lorentz violation
can be calculated in perturbation theory.
For muon propagation with momentum $\pvec$,
the perturbative hamiltonian $\de h(\pvec)$
arising from Lorentz-violating operators of arbitrary mass dimension
is derived in Ref.\ \cite{km13},
and the motion in the classical limit
follows a geodesic in a pseudo-Finsler spacetime
\cite{finsler1,finsler2}.
For experimental applications,
it is convenient to adopt a decomposition 
of the hamiltonian using spherical coordinates,
which reveals that the perturbative terms 
are controlled by eight sets of spherical coefficients 
for Lorentz violation.
These are denoted as
$\acoef{d}{njm}$, 
$\ccoef{d}{njm}$,
$\gzBcoef{d}{njm}$, 
$\goBcoef{d}{njm}$,
$\goEcoef{d}{njm}$,
$\HzBcoef{d}{njm}$,
$\HoBcoef{d}{njm}$,
$\HoEcoef{d}{njm}$,
where $d$ is the mass dimension of the corresponding operator
and the allowed ranges of the indices $n$, $j$, $m$
are given in Table III of Ref.\ \cite{km13}. 
The $g$- and $H$-type coefficients
are associated with spin operators 
causing birefringence of the muon propagation,
which can be interpreted 
as a Larmor-like precession of the muon spin $\Svec$
and affects the spin-precession frequencies $\om_s^\pm$.

Denoting the corresponding pieces of $\de h(\pvec)$ as
$h_g = \mbf{h}_g\cdot \mbf\si$ 
and $h_H = \mbf{h}_H\cdot \mbf\si$,
the rate of change of the spin expectation value
for the $\mu^-$ 
due to Lorentz violation is given by
\cite{km13}
\beq
\fr{d\vev{\Svec}}{dt} 
\approx 2(\mbf{h}_g+\mbf{h}_H) \times\vev{\Svec} .
\label{sppr}
\eeq
The correction to the muon spin-precession frequency 
can then be identified as
$\de\mbf{\om}_s^- = 2(\mbf{h}_g + \mbf{h}_H)$.
The result for the antimuon $\mu^+$
follows by changing the sign of the $g$-type coefficients,
which control CPT-odd operators in $\de h$. 
Since the cyclotron frequency by definition
is produced by level shifts proportional to the magnetic field $B$,
which is tiny in natural units
(1 T $\simeq 2\times 10^{-16}$ GeV$^2$),
Lorentz-violating corrections to this frequency 
are determined by the product of two small quantities
and hence can be neglected.
The corrections to the $\mu^\pm$ anomaly frequencies are therefore given by 
\beq
\de\mbf{\om}_a^\pm = \pm 2 \mbf{h}_g + 2 \mbf{h}_H .
\label{shiftaf}
\eeq
In experimental applications,
the detectors lie in the plane of the storage ring
and so only the perpendicular component of $\de\mbf{\om}_a^\pm$
is measured.
Moreover,
only orbital averages are observed,
so the couplings involving both Lorentz violation and the muon momentum
can contribute only when cylindrically symmetric
about the vertical axis through the storage ring.

The result \rf{shiftaf} holds in the laboratory frame.
As discussed in Sec.\ \ref{Basics} in the context of muonic bound states,
the rotation of the Earth
induces time dependence of some coefficients in the laboratory frame.
Disregarding for the moment
effects from the revolution of the Earth about the Sun,
which are suppressed by about four orders of magnitude,
the time dependence of a generic coefficient 
$\K^{\rm lab}_{jm}$
in the frame of a laboratory
with $x$ axis pointing south and $y$ axis pointing east 
is given by
\cite{km09}
\beq
\K^{\rm lab}_{jm}=
\sum_{m^\prime}e^{im^\prime \om_\oplus T_\oplus}
d^{(j)}_{m m^\prime}(-\chi)
\K^{\rm Sun}_{jm^\prime}
\eeq
in terms of the corresponding coefficients
$\K^{\rm Sun}_{jm}$
in the canonical Sun-centered frame.
As before,
$\om_\oplus \simeq 2\pi/(23 {\rm ~h} ~56 {\rm ~m})$
is the Earth's sidereal frequency 
and $T_\oplus$ is the sidereal time,
while the little Wigner matrices $d^{(j)}_{m m^\prime}$ 
are taken as defined in Eq.\ (136) of Ref.\ \cite{km09}
and $\ch$ is the colatitude of the experiment in the northern hemisphere.
In what follows,
we adopt the values  
$\ch \simeq 43.7^\circ$ at CERN,
$\ch \simeq 49.1^\circ$ at BNL,
$\ch \simeq 48.2^\circ$ at Fermilab,
and $\ch \simeq 53.5^\circ$ at J-PARC.

Combining the above results
yields the experimentally observable perturbative shift 
$\de\om_a^\pm$
of the anomaly frequency due to Lorentz violation,
expressed in terms of spherical coefficients in the Sun-centered frame.
The result is
\beq
\de\om_a^\pm = 
2\sum_{dnjm} E_0^{d-3} 
e^{im \om_\oplus T_\oplus} G_{jm}(\ch) 
\big(\hh^{(d)}_{njm} \pm \hg^{(d)}_{njm}\big),
\label{omega-sun-frame}
\eeq
where $E_0$ is the unperturbed muon energy.
This is a central result for studying Lorentz and CPT violation
via experiments measuring the muon anomalous magnetic moment.

In Eq.\ \rf{omega-sun-frame},
the dimensionless factor 
\beq
G_{jm}(\ch) \equiv 
\sqrt{j(j+1)}~\syjm 1 {j0}(\pi/2,0) d^{(j)}_{0m}(-\ch)
\label{geom}
\eeq
is purely geometrical 
and involves the spin-weighted spherical harmonics $\syjm 1 {j0}(\th,\ph)$
of spin weight 1
defined according to Appendix A of Ref.\ \cite{km09}.
The contribution to $\de\om_a^\pm$ vanishes for even $j$
because $\syjm 1 {j0}(\pi/2,0)$ does,
while for odd $j=2k+1$ we have 
\beq
\syjm 1 {(2k+1)0}(\pi/2,0) =
\fr{(-1)^k (2k-1)!!}{2^{k+1} k!}
\sqrt{ \fr{(1+2k)(3+4k)}{2\pi (1+k)}} .
\eeq
Table \ref{gjmvalues} 
lists some numerical values
of the factor $G_{jm}(\ch)$ 
relevant for the CERN, BNL, Fermilab, and J-PARC experiments.

\begin{table}
\centering
\caption{Some useful values of $G_{jm}(\chi)$
for the CERN, BNL, Fermilab, and J-PARC experiments.}
\renewcommand\arraystretch{1.5}
\setlength{\tabcolsep}{7pt}
\begin{tabular}{cccccc}
\noalign{\smallskip}
\hline
\hline
$j$ & $m$ & CERN & BNL & Fermilab & J-PARC\\
\hline
1&  0 & $0.353$ & $0.320$ & $0.326$ & $0.291$
\\
&  $\pm 1$ & $\mp 0.239$ & $\mp 0.261$ & $\mp 0.258$ & $\mp 0.278$
\\
3&  0 & $0.156$ & $0.314$ & $0.291$ & $0.410$
\\
&  $\pm 1$ & $\pm 0.540$ & $\pm 0.419$  & $\pm 0.441$ & $\pm 0.300$
\\
&  $\pm 2$ & $-0.529$ & $-0.573$  & $-0.568$ & $-0.589$
\\
&  $\pm 3$ & $\pm 0.206$ & $\pm 0.270$  & $\pm 0.259$ & $\pm 0.325$
\\
5& 0 & $-0.694$ & $-0.493$ & $-0.536$ & $-0.245$
\\
&  $\pm 1$ & $\pm 0.241$ & $\pm 0.518$  & $\pm 0.481$ & $\pm 0.639$
\\
&  $\pm 2$ & $0.623$ & $0.340$  & $0.392$ & $0.0750$
\\
&  $\pm 3$ & $\mp 0.792$ & $\mp 0.801$  & $\mp 0.806$ & $\mp 0.736$
\\
&  $\pm 4$ & $0.453$ & $0.588$  & $0.566$ & $0.684$
\\
&  $\pm 5$ & $\mp 0.137$ & $\mp 0.215$  & $\mp 0.200$ & $\mp 0.292$
\\
7& 0 & $-0.170$ & $-0.634$ & $-0.576$ & $-0.773$
\\
\hline
\hline
\end{tabular}
\label{gjmvalues}
\end{table}

The h\' a\v cek coefficients 
$\hg^{(d)}_{njm}$, $\hh^{(d)}_{njm}$
appearing in the expression \rf{omega-sun-frame}
represent the linear combinations of spherical coefficients
that are observable in the experiments.
They are defined as 
\bea
\hg^{(d)}_{njm} &\equiv& \be^n g^{(d)(0B)}_{njm} 
+ \sqrt{\frac{2}{j(j+1)}} \be^{n+2} g^{(d)(1B)}_{(n+2)jm},
\nonumber\\
\hh^{(d)}_{njm} &\equiv& \be^n H^{(d)(0B)}_{njm} 
+ \sqrt{\frac{2}{j(j+1)}} \be^{n+2} H^{(d)(1B)}_{(n+2)jm},
\quad
\label{defs}
\eea
where the muon velocity is $\be = \sqrt{1-1/\ga^2}$ as usual.
Note that all coefficients with $m\neq 0$ are complex,
and coefficients with negative $m$
are related to those with positive $m$ via expressions of the form
$\K{}^*_{jm} = (-1)^m \K_{j(-m)}$
\cite{km13}.
Also,
only coefficients with even $n$ contribute to Eq.\ \rf{defs},
as those with odd $n$ come only with even $j$
and hence cancel via the geometrical factor \rf{geom}.
For example,
we find 32 independent observable combinations 
can contribute for $d\leq 6$,
and they are denoted as
$\hh^{(3)}_{01m}$,
$\hg^{(4)}_{01m}$,
$\hh^{(5)}_{01m}$,
$\hh^{(5)}_{21m}$,
$\hh^{(5)}_{23m}$,
$\hg^{(6)}_{01m}$,
$\hg^{(6)}_{21m}$,
and $\hg^{(6)}_{23m}$.
Moreover,
the coefficients on the right-hand side of Eq.\ \rf{defs}
are understood to contribute only if their indices
lie in the ranges given in Table III of Ref.\ \cite{km13}.
For example,
for $d=3$ 
the coefficient $\hh^{(3)}_{njm}$
contains only $ H^{(3)(0B)}_{njm}$ 
because $H^{(d)(1B)}_{njm}$ exists only for $d\geq 5$.
Along similar lines,
$\hh^{(5)}_{2jm}$
contains only $ H^{(5)(0B)}_{2jm}$ 
because $H^{(5)(1B)}_{njm}$ exists only for $n\leq 2$.

The expression \rf{omega-sun-frame} for $\de\om_a^\pm$ 
encompasses effects from Lorentz-violating operators 
of arbitrary mass dimensions. 
In the appropriate limit,
it reduces to the analogous result (11) for $\de\om_a^+$
derived in Ref.\ \cite{bkl} 
in terms of the minimal-SME cartesian coefficients
$b_\mu$, $d_{\mn}$, and $H_{\mn}$.
One set of predicted effects for the general case includes 
sidereal variations of $\de\om_a^{\pm}$
at harmonics of $\om_\oplus$.
Another prediction is a difference 
$\De\om_a = \de\om_a^{+} - \de\om_a^{-}$
between the anomaly frequencies of the muon and antimuon,
with both constant and time-varying components.
These predictions are discussed in the next two subsections.
In addition,
the orbital motion of the Earth about the Sun
introduces further sensitivities to Lorentz violation
beyond those in Eq.\ \rf{omega-sun-frame}.
Treating these requires a separate analysis, 
which is the subject of Sec.\ \ref{Annual variations}.

The form of the correction $\de\om_a^\pm$
given by Eq.\ \rf{omega-sun-frame} 
reveals that sensitivity to coefficients of larger $d$
typically increases with the particle energy.
This behavior places the planned experiments
at Fermilab
\cite{fermilab} 
and J-PARC
\cite{j-parc}
in distinct positions,
as it indicates that each enjoys different sensitivities
to some coefficient combinations.
Both have a similar overall potential reach 
for minimal-SME spherical coefficients,
but the smaller value of $\ga$ to be used at J-PARC
implies an improvement of an order of magnitude
in sensitivity to certain minimal-SME cartesian coefficients such as $b_3$.
In contrast,
the higher-energy particles to be used at Fermilab
lead to greater sensitivity to nonminimal coefficients,
and the design of the Fermilab experiment may ultimately
permit measurements with both muons and antimuons, 
a feature unavailable to the currently proposed J-PARC setup.
Moreover,
the differing colatitudes and energies 
of the Fermilab and J-PARC experiments
suggests that combining results for antimuons
would permit constraints on coefficient combinations
inaccessible to any single experiment.

\subsection{Muon-antimuon comparison}
\label{Muon-antimuon comparison}

\subsubsection{CPT-odd effects}
\label{CPT-odd effects}

When CPT violation is present in the muon sector,
differences between the anomaly frequencies $\om_a^+$ and $\om_a^-$
can appear.
Simultaneous measurement of the two frequencies is experimentally infeasible,
but a comparison of them averaged over many sidereal days
can directly isolate the CPT violation.
Denoting the time-averaged anomaly-frequency difference
by $\vev{\De\om_a}$,
we obtain
\bea
\vev{\De\om_a} &=& 
\vev{\de\om^{+}_a} - \vev{\de\om^{-}_a} 
\nonumber\\
&=& 4 \sum_{dnj} E_0^{d-3} G_{j0}(\ch) 
\hg^{(d)}_{nj0}.
\label{deoma}
\eea
This approach restricts attention to 
coefficients $\hg^{(d)}_{nj0}$ for CPT violation 
having index $m=0$,
which control azimuthally isotropic operators in the Sun-centered frame
and hence exhibit no sidereal variations at leading order.

The BNL experiment 
\cite{bnl08}
reported a measurement of the figure of merit 
$\vev{\De\om_a}/m_\mu$,
which using $m_\mu = 105.7$ MeV
gives 
the constraint
\beq
\sum_{dnj} E_0^{d-3} G_{j0}(\ch) \hg^{(d)}_{nj0}
= -(2.3\pm 2.4) \times 10^{-25} {\rm ~GeV}.
\label{omaconst}
\eeq
Paralleling the above discussion of muonic atoms,
it is useful to extract from this expression
the attained sensitivities to individual spherical coefficients,
taking only one coefficient to be nonzero at a time.
The resulting constraints 
are compiled in Table \ref{table:muonantimuon2}
for $d\leq 10$.
Note that these values correspond closely to  
limits on the coefficients $g^{(d)(0B)}_{nj0}$ 
derived using the definition \rf{defs}
because $\be\simeq 1$ to an excellent approximation,
while limits on the coefficients $g^{(d)(1B)}_{nj0}$
can be obtained by scaling with a factor of $\sqrt{j(j+1)/2}$.

\begin{table}
\centering
\caption{Constraints on spherical coefficients 
determined from the anomaly-frequency difference in the BNL experiment. 
Units are GeV$^{4-d}$.} 
\renewcommand\arraystretch{1.7}
\setlength{\tabcolsep}{7pt}
\begin{tabular}{c c c}
\noalign{\smallskip}
\hline
\hline
$d$ & Coefficient & Constraint\\
\hline  
4 & $\hg^{(4)}_{010}$ & 
$(-2.3\pm 2.4)\times 10^{-25}$ \\ 
6 & $\hg^{(6)}_{010}$, $\hg^{(6)}_{210}$ & 
$(-2.4\pm 2.5)\times 10^{-26}$ \\  
& $\hg^{(6)}_{230}$ & 
$(-2.5\pm 2.5)\times 10^{-26}$ \\  
8 & $\hg^{(8)}_{010}$, $\hg^{(8)}_{210}$, $\hg^{(8)}_{410}$ & 
$(-2.5\pm 2.6)\times 10^{-27}$ \\  
& $\hg^{(8)}_{230}$, $\hg^{(8)}_{430}$ & 
$(-2.6\pm 2.6)\times 10^{-27}$ \\  
& $\hg^{(8)}_{450}$ & 
$(1.6\pm 1.7)\times 10^{-27}$ \\  
10 & $\hg^{(10)}_{010}$, $\hg^{(10)}_{210}$, 
$\hg^{(10)}_{410}$, $\hg^{(10)}_{610}$ & 
$(-2.6\pm 2.7)\times 10^{-28}$ \\  
& $\hg^{(10)}_{230}$, $\hg^{(10)}_{430}$, $\hg^{(10)}_{630}$ & 
$(-2.7\pm 2.7)\times 10^{-28}$ \\  
& $\hg^{(10)}_{450}$, $\hg^{(10)}_{650}$ & 
$(1.7\pm 1.7)\times 10^{-28}$ \\  
& $\hg^{(10)}_{670}$ & 
$(1.3\pm 1.4)\times 10^{-28}$ \\ 
\hline
\hline
\end{tabular}
\label{table:muonantimuon2}
\end{table}

Results for the minimal SME are given 
as a limiting case of the above.
Indeed,
the minimal-SME coefficient $\hg^{(4)}_{010}$ 
can be expressed in terms of cartesian coefficients 
in the Sun-centered frame as
\beq
\hg^{(4)}_{010} = 
\fr 1{\ga E_0}
\sqrt{\fr{4\pi}{3}} 
\big( b_Z - m_\mu g^{({\rm A})}_Z
+ (1+\frac 3 2\be^2 \ga^2) m_\mu g^{({\rm M})}_{XYT}\big),
\label{bound-b}
\eeq
where the superscripts (A) and (M)
denote the irreducible axial and irreducible mixed-symmetry
combinations of the coefficients $g^{}_{\ka\la\nu}$,
respectively
\cite{krt08,fittante}. 
This result extends the one derived 
in the original theoretical treatment 
\cite{bkl},
which excludes the coefficients $g^{}_{\ka\la\nu}$ 
on the grounds of an expected suppression
relative to other minimal-SME coefficients
arising from the breaking of SU(2)$\times$U(1) symmetry
\cite{ck}.
It thereby reveals that the measurement of $b_Z$
reported by the BNL experiment
\cite{bnl08}
can be extended to
\bea
b_Z - m_\mu g^{({\rm A})}_Z
+ (1+\frac 32 \be^2\ga^2) m_\mu g^{({\rm M})}_{XYT}
&&
\nonumber\\
&&
\hskip -80pt
= -(1.0\pm 1.1)\times 10^{-23} {\rm ~GeV}.
\qquad
\eea
This represents the first reported constraint
containing muon-sector $g$-type coefficients in the minimal SME.
The comparatively large boost in this experiment
provides an enhanced sensitivity to the mixed-symmetry combination
taken by itself,
$m_\mu g^{({\rm M})}_{XYT} = -(7.8 \pm 8.5)\times 10^{-27}$ GeV,
that compares favorably with the Planck-suppressed
ratio $m_\mu^2/M_P\simeq 9.2 \times 10^{-22}$ GeV.
 
The existing proposals for 
the forthcoming Fermilab
\cite{fermilab}
and J-PARC
\cite{j-parc}
experiments
are focused on measurements of the antimuon anomalous magnetic moment.
However,
if future studies of muons could also be performed 
at similar sensitivities,
then the constraints given in Table \ref{table:muonantimuon2}
could be improved by a factor of roughly 5 
from precision alone.

\subsubsection{CPT-even effects}
\label{CPT-even effects}

The availability of experiments at different latitudes
also offers access 
to isotropic coefficients for CPT-even Lorentz violation
\cite{bnl08}.
The idea is that the predicted 
time-averaged antimuon anomaly frequency $\om_a^+(\ch_1)$ 
at colatitude $\ch_1$ 
differs from the time-averaged muon anomaly frequency $\om_a^-(\ch_2)$
at colatitude $\ch_2\neq\ch_1$,
and this difference is sensitive also to CPT-even effects.
Since the two experiments typically also have distinct magnetic fields,
it is useful to work in terms of the ratios
$\R^\pm = \om_a^\pm/\om_p$ 
of the muon anomaly frequencies to the proton cyclotron frequency,
which removes the dependence on the magnetic field
and is widely used in experimental analyses.

The difference $\vev{\De\R (1,2)}$
between the time-averaged values of $\R^+ (\ch_1)$ and $\R^-(\ch_2)$
is found from Eq.\ \rf{omega-sun-frame} to be
\bea
\vev{\De\R(1,2)}&\equiv&
\vev{\R^+(\ch_1)}-\vev{\R^-(\ch_2)}
\nonumber\\
&=& 
2\sum_{dnj} E_0^{d-3} 
\left( \fr{G_{j0}(\ch_1)} {\om_p(\ch_1)} 
+ \fr{G_{j0}(\ch_2)} {\om_p(\ch_2)} \right) \hg^{(d)}_{nj0}
\nonumber\\
&&
\hskip -10pt
+2\sum_{dnj} E_0^{d-3} 
\left( \fr{G_{j0}(\ch_1)} {\om_p(\ch_1)} 
- \fr{G_{j0}(\ch_2)} {\om_p(\ch_2)} \right) \hh^{(d)}_{nj0}.
\nonumber\\
\label{diffchi}
\eea
As can be seen from Table \ref{gjmvalues},
the dimensionless factor $G_{j0}(\ch_1) - G_{j0}(\ch_2)$ is small, 
so the sensitivity to the coefficients $\hh^{(d)}_{nj0}$
is reduced compared with that to CPT-odd effects.
For practical purposes,
we can therefore assume here that the coefficients
$\hg^{(d)}_{nj0}$
have been excluded at a sufficient precision
so that attention can be focused purely on CPT-even effects.
Then,
only the piece of Eq.\ \rf{diffchi}
involving the coefficients $\hh^{(d)}_{nj0}$ contributes.
Note that these coefficients
carry index $m=0$ and therefore cannot be detected via sidereal variations,
so an analysis using Eq.\ \rf{diffchi}
offers an interesting avenue for exploration
of CPT-even effects that otherwise might escape detection.

The CERN
\cite{cern}
and BNL 
\cite{bnl}
experiments
have each reported values of $\R^\pm$
and are located at colatitudes differing by about 5$^\circ$.
These results can be used to calculate 
$\De\R(\rm{CERN, BNL})$,
for example,
which involves antimuons at CERN and muons at BNL.
The CERN experiment at $\ch \simeq 43.7^\circ$ 
measured
$\R^+ = 3.707173(36) \times 10^{-3}$ 
with $\om_p/2\pi \simeq 6.278302(5) \times 10^7$ Hz,
while
the BNL experiment at $\ch \simeq 49.1^\circ$ 
obtained
$\R^- = 3.7072083(26) \times 10^{-3}$
with 
$\om_p/2\pi \simeq 6.1791400(11) \times 10^7$ Hz.
Using these values, 
we find
$\De\R(\rm{CERN, BNL}) = (-3.5 \pm 3.6) \times 10^{-8}$.
With this value and neglecting as comparatively small 
the contributions from $\hg^{(d)}_{nj0}$,
we obtain the bound
\bea
\sum_{dnj} E_0^{d-3} 
\left( \fr{G_{j0}(\ch_1)} {\om_p(\ch_1)} 
- \fr{G_{j0}(\ch_2)} {\om_p(\ch_2)} \right) \hh^{(d)}_{nj0}
&&
\nonumber\\
&&
\hskip -80 pt
< (-1.8 \pm 1.8) \times 10^{-8}.
\label{diffchilimit}
\eea
As before,
we can gain insight by extracting from this bound
the attained sensitivities 
to each individual coefficient at a time.
The results for $d\leq 9$ are displayed in
Table \ref{hcoeffresults}.
Additional constraints can be obtained by calculating
$\De\R(\rm{BNL, CERN})$,
which instead involves 
muons at CERN and antimuons at BNL.
This gives a comparable sensitivity of
$\De\R(\rm{BNL, CERN}) = (-5.1 \pm 3.7) \times 10^{-8}$
and slightly weaker constraints on the individual coefficients.
Improved results along these lines
can be expected once measurements have been made
by the forthcoming Fermilab and J-PARC experiments. 

\begin{table}
\centering
\caption{Constraints on spherical coefficients
determined from the difference between 
CERN antimuon and BNL muon anomaly frequencies.
Units are GeV$^{4-d}$.}
\renewcommand\arraystretch{1.7}
\setlength{\tabcolsep}{7pt}
\begin{tabular}{c c c}
\noalign{\smallskip}
\hline
\hline
$d$ & Coefficient & Constraint\\
\hline
3 & $\hh^{(3)}_{010}$ &
$(-1.6\pm 1.7)\times 10^{-22}$ \\
5 & $\hh^{(5)}_{010}$, $\hh^{(5)}_{210}$ &
$(-1.7\pm 1.7)\times 10^{-23}$ \\
& $\hh^{(5)}_{230}$ &
$(2.9\pm 3.0)\times 10^{-24}$ \\
7 & $\hh^{(7)}_{010}$, $\hh^{(7)}_{210}$, $\hh^{(7)}_{410}$ &
$(-1.7\pm 1.8)\times 10^{-24}$ \\
& $\hh^{(7)}_{230}$, $\hh^{(7)}_{430}$ &
$(3.0\pm 3.1)\times 10^{-25}$ \\
& $\hh^{(7)}_{450}$ &
$(2.6\pm 2.6)\times 10^{-25}$ \\
9 & $\hh^{(9)}_{010}$, $\hh^{(9)}_{210}$,
$\hh^{(9)}_{410}$, $\hh^{(9)}_{610}$ &
$(-1.8\pm 1.9)\times 10^{-25}$ \\
& $\hh^{(9)}_{230}$, $\hh^{(9)}_{430}$, $\hh^{(9)}_{630}$ &
$(3.2\pm 3.3)\times 10^{-26}$ \\
& $\hh^{(9)}_{450}$, $\hh^{(9)}_{650}$ &
$(2.7\pm 2.7)\times 10^{-26}$ \\
& $\hh^{(9)}_{670}$ &
$(-1.1\pm 1.1)\times 10^{-26}$ \\
\hline
\hline
\end{tabular}
\label{hcoeffresults}
\end{table}

\subsection{Sidereal variations}
\label{Sidereal variations}

The general expression \rf{omega-sun-frame} 
for $\de\om_a^{\pm}$ shows that 
nonzero coefficients for Lorentz violation with $m\neq 0$
lead to the variation of $\om_a^{\pm}$
with sidereal time $T_\oplus$. 
The variation is a superposition of oscillations
of different frequencies.
A given term in the sum has harmonic frequency $m\om_\oplus$,
where $m$ is the index on the corresponding coefficient.
The total amplitude of the $m$th harmonic is 
\beq
A^{\pm}_{m} = 
\bigg|4 \sum_{dnj} E_0^{d-3} G_{jm}(\chi) 
\big[\hh^{(d)}_{njm} \pm E_0\, \hg^{(d+1)}_{njm}\big]
\bigg|,
\quad m\neq 0.
\label{sid-var-coeff}
\eeq
In evaluating this expression,
recall that the argument of the modulus is complex in general
because for $m\neq 0$ 
the coefficients can have real and imaginary parts.

The above amplitude is valid for any finite range of 
the operator mass dimension $d$.
The maximum value of $d$ in this range
determines the highest harmonic variation appearing in the signal.
To illustrate this,
focus on one specific value of $d$ at a time,
as is typical in data analyses 
studying the effects of nonminimal Lorentz-violating operators
\cite{tables}.
Since the specific value of $d$ 
determines the corresponding range of the coefficient indices $n$ and $j$
\cite{km13},
which in turn controls the largest possible size of $m$,
it follows that the allowed harmonics for a given $d$
include values of $m$ up to a definite maximum $m_{\rm max}$.
For odd $d$ 
we find 
$m_{\rm max} = d-2$,
while for even $d$ we obtain  
$m_{\rm max} = d-3$.
For example,
when $d=3$ and 4 
only the fundamental sidereal frequency $\om_\oplus$ appears,
while $d=5$ and 6 operators 
are accompanied also by variations 
at the frequencies $2\om_\oplus$ and $3\om_\oplus$. 
 
The E821 experiment at BNL searched for sidereal variations
at the fundamental sidereal frequency $\om_\oplus$
\cite{bnl08},
obtaining the bounds 
$A^{+}_1 \le 2.1 \times 10^{-24}\,{\rm GeV}$ 
and 
$A^{-}_1 \le 4.0 \times 10^{-24}\,{\rm GeV}$
at the 95\% confidence level
and placing the tightest limits to date 
on minimal-SME muon-sector coefficients.
In the present context,
Eq.\ \rf{sid-var-coeff} reveals that these results 
also bound some combinations of spherical coefficients
associated with Lorentz-violating operators 
of arbitrarily high mass dimension.
We can illustrate this explicitly
by determining the sensitivities to individual spherical coefficients
for a range of values of $d$,
with only the real or imaginary part of a single coefficient 
assumed nonzero at a time.
Table \ref{sidconst} lists the resulting constraints
on all spherical coefficients with $m=1$ and $d\leq 8$.
A reanalysis of the BNL data constraining higher harmonics
could yield measurements on all the remaining coefficients 
with $m\neq 0$ as well.
Indeed,
the Lomb power spectrum displayed in Fig.\ 2 of Ref.\ \cite{bnl08}
suggests no varying signal at the various sidereal harmonics,
offering the potential for tight bounds on these coefficients.

\begin{table}
\centering
\caption{Constraints on the moduli of the real and imaginary parts 
of spherical coefficients determined from sidereal variations 
of the antimuon anomaly frequency in the BNL experiment.
Units are GeV$^{4-d}$.}
\setlength{\tabcolsep}{9pt}
\renewcommand{\arraystretch}{1.5}
\begin{tabular}{ccc }
\noalign{\smallskip}
\hline\hline
$d$ & Coefficient & Constraint on\\ 
[-4pt]
& $\hk$& $|\Re\hk|$, $|\Im\hk|$\\ 
\hline  
3 & $\hh^{(3)}_{011}$ & 
$<2.0 \times 10^{-24}$
\\ 
4 & $\hg^{(4)}_{011}$ & 
$<6.6 \times 10^{-25}$
\\
5 & $\hh^{(5)}_{011}$, $\hh^{(5)}_{211}$ & 
$<2.1 \times 10^{-25}$
\\
& $\hh^{(5)}_{231}$ & 
$<1.3 \times 10^{-25}$ 
\\ 
6 & $\hg^{(6)}_{011}$, $\hg^{(6)}_{211}$ & 
$<6.8 \times 10^{-26}$ 
\\
& $\hg^{(6)}_{231}$ & 
$<4.3 \times 10^{-26}$ 
\\
7 & $\hh^{(7)}_{011}$, $\hh^{(7)}_{211}$, $\hh^{(7)}_{411}$ & 
$<2.2 \times 10^{-26}$ 
\\
& $\hh^{(7)}_{231}$, $\hh^{(7)}_{431}$ & 
$<1.4 \times 10^{-26}$ 
\\
& $\hh^{(7)}_{451}$ & 
$<1.1 \times 10^{-26}$
\\
8 & $\hg^{(8)}_{011}$, $\hg^{(8)}_{211}$, $\hg^{(8)}_{411}$ & 
$<7.1 \times 10^{-27}$ 
\\
& $\hg^{(8)}_{231}$, $\hg^{(8)}_{431}$ & 
$<4.5 \times 10^{-27}$ 
\\
& $\hg^{(8)}_{451}$ & 
$<3.6 \times 10^{-27}$ 
\\
\hline\hline
\end{tabular}
\label{sidconst}
\end{table}

In the limiting case of the minimal SME,
the amplitudes \rf{sid-var-coeff}
can be expressed in terms of cartesian coefficients
for operators of dimension $d=3$ and 4 using the relationships
\bea
\Re \hh^{(3)}_{011} \pm E_0 \Re\hg^{(4)}_{011}
&=& -\sqrt{\frac{2\pi}{3}} ~\hb{}^\pm_X,
\nonumber\\
\Im \hh^{(3)}_{011} \pm E_0 \Im\hg^{(4)}_{011}
&=& -\sqrt{\frac{2\pi}{3}} ~\hb{}^\pm_Y
\eea
in the Sun-centered frame.
Here,
the combinations 
\bea
\hb{}^\pm_J &\equiv & 
\pm \fr 1\ga \big( b^{}_J - m_\mu g^{({\rm A})}_J\big)
+ \half \ep_{JKL} H^{}_{KL} +  m_\mu d^{}_{JT}
\nonumber\\
&& 
\pm \fr{1}{2\ga} (1+\frac 32\be^2 \ga^2 ) m_\mu \ep_{JKL}g^{({\rm M})}_{KLT} 
\label{bhacek}
\eea
generalize the quantities introduced in Ref.\ \cite{bkl}
to include also contributions from 
the antisymmetric and mixed-symmetry irreducible combinations
of the $g^{}_{\ka\la\nu}$ coefficients,
in analogy with Eq.\ \rf{bound-b}.
This shows that the two bounds reported as Eq.\ (11) of Ref.\ \cite{bnl08}
incorporate also sensitivity to the $g$-type coefficients.
For example,
in a model with only $g^{({\rm M})}_{KLT}$ nonzero,
the constraint
$m_\mu \sqrt{(g^{({\rm M})}_{XZT})^2 + (g^{({\rm M})}_{YZT})^2}
<1.1\times 10^{-27}$ GeV 
at the 95\% confidence level is obtained.
We remark in passing
that in the nonrelativistic limit $\be\to 0$, $\ga\to 1$
the h\'a\v cek coefficients \rf{bhacek}
reduce to combinations of the standard cartesian tilde coefficients
\cite{tables},
yielding the correspondences 
$\hb{}^+_J \to \widetilde b{}^*_J$,
$\hb{}^-_J \to - \widetilde b_J$.

The constraints in Table \ref{sidconst}
are a consequence 
of the 0.54 ppm precision attained by the BNL experiment
\cite{bnl}.
Future proposals aim to achieve 
0.14 ppm at Fermilab 
\cite{fermilab}
and 0.1 ppm at J-PARC
\cite{j-parc},
which would offer the opportunity to sharpen the values 
in Table \ref{sidconst} 
by about a factor of five.
The energy dependence of the amplitudes \rf{sid-var-coeff}
suggests that the Fermilab and J-PARC experiments 
will achieve approximately the same sidereal reach 
to the spherical coefficients with $d=3$, 
with the latter's sensitivity for $d\geq 4$
suppressed by a factor of about $10^{d-3}$. 
However,
the J-PARC experiment enjoys
an additional improvement of a factor of about 10
in the sensitivity to certain coefficients 
that are accompanied by factors of $1/\ga$,
such as $b_J$ in Eq.\ \rf{bhacek}.
More broadly,
comparing results from experiments at different $\ga$
offers the opportunity to disentangle coefficients,
as exemplified by the structure of Eq.\ \rf{bhacek}.

\subsection{Annual variations}
\label{Annual variations}

\begin{table*}
\centering
\caption{Factors forming the expansion of 
the muon and antimuon observables
in the Sun-centered frame.
For each particle,
the complete expression is obtained by 
multiplying the factors in each row
and adding all the relevant rows.}  
\setlength{\tabcolsep}{3pt}
\renewcommand{\arraystretch}{1.5}
\begin{tabular}{ccccc}
\hline\hline  
& Boost & Sidereal & Colatitude & Coefficient \\ 
[-4pt]
Particle & factor & factor & factor & factor \\ 
\hline
$\mu^-$ & 1 & 1 & $\cc$ & 
$\hb_Z$ \\
& 1 & $\codt$ & $\sc$ & 
$\hb_X$ \\
& 1 & $\sodt$ & $\sc$ & 
$\hb_Y$ \\
& $\be_\oplus$ & $\cto$ & $\cc$ & 
$- \ce (\hH_{TX}) +\se (-\hg_T +2\hd_+ -\hd_Q )$ \\
& $\be_\oplus$ & $\sto$ & $\cc$ & 
$-\hd_{ZX} - \hH_{TY}$ \\
& $\be_\oplus$ & $\codt \cto$ & $\sc$ & 
$\ce (\hd_{XY} +\hH_{TZ}) -\se \hH_{TY}$ \\
& $\be_\oplus$ & $\codt \sto$ & $\sc$ & 
$-\half\hb_T -\half\hd_- +\hg_c +\frac{3}{2}\hg_T -2\hd_+ +\half\hd_Q$ \\
& $\be_\oplus$ & $\sodt \cto$ & $\sc$ & 
$\ce ( -\half\hb_T -\half\hd_- +\hg_c -\half\hg_T +2\hd_+ -\half\hd_Q )
+\se(\hd_{YZ} + \hH_{TX})$ \\
& $\be_\oplus$ & $\sodt \sto$ & $\sc$ & 
$\hH_{TZ}$ \\
& $\be_L$ & 1 & $\sc$ & 
$-\half \hd_{XY} -\hH_{TZ}$ \\
& $\be_L$ & $\codt$ & $\cc$ & 
$\hH_{TX}$ \\
& $\be_L$ & $\sodt$ & $\cc$ & 
$\hd_{ZX} +\hH_{TY}$ \\
& $\be_L$ & $\ctodt$ & $\sc$ & 
$-\half \hd_{XY}$ \\
& $\be_L$ & $\stodt$ & $\sc$ & 
$\half \hb_T +\half \hd_- -\hg_c -\half\hg_T$ \\
[8pt]
$\mu^+$ & 1 & 1 & $\cc$ & 
$\hb{}^*_Z$ \\
& 1 & $\codt$ & $\sc$ & 
$\hb{}^*_X$ \\
& 1 & $\sodt$ & $\sc$ & 
$\hb{}^*_Y$ \\
& $\be_\oplus$ & $\cto$ & $\cc$ & 
$\ce( -2 \hg_{XY} +\hH_{TX}) +\se( - 2 \hb_T + \hg_T -2\hd_+ +\hd_Q )$ \\
& $\be_\oplus$ & $\sto$ & $\cc$ & 
$\hd_{ZX} -2 \hg_{YX} + \hH_{TY}$ \\
& $\be_\oplus$ & $\codt \cto$ & $\sc$ & 
$\ce ( -\hd_{XY} + 2 \hg_{ZY} -\hH_{TZ}) +\se (-2 \hg_{YZ} + \hH_{TY})$ \\
& $\be_\oplus$ & $\codt \sto$ & $\sc$ & 
$\half\hb_T +\half\hd_- +\hg_c +\half\hg_T +2\hd_+ -\half\hd_Q$ \\
& $\be_\oplus$ & $\sodt \cto$ & $\sc$ & 
$\ce( -\frac 32 \hb_T +\half\hd_- +\hg_c +\half\hg_T -2\hd_+ +\half\hd_Q )
+\se(- \hd_{YZ} + 2 \hg_{XZ} - \hH_{TX})$ \\
& $\be_\oplus$ & $\sodt \sto$ & $\sc$ & 
$2 \hg_{ZX} - \hH_{TZ}$ \\
& $\be_L$ & 1 & $\sc$ & 
$\half \hd_{XY} -\hg_{ZX} -\hg_{ZY} +\hH_{TZ}$ \\
& $\be_L$ & $\codt$ & $\cc$ & 
$2 \hg_{XY} - \hH_{TX}$ \\
& $\be_L$ & $\sodt$ & $\cc$ & 
$- \hd_{ZX} +2\hg_{YX} -\hH_{TY}$ \\
& $\be_L$ & $\ctodt$ & $\sc$ & 
$\half \hd_{XY} + \hg_{ZX} - \hg_{ZY} $ \\
& $\be_L$ & $\stodt$ & $\sc$ & 
$\half \hb_T -\half \hd_- -\hg_c -\half\hg_T$ \\
\hline\hline
\end{tabular}
\label{orbitalmuon}
\end{table*}

\begin{table*}
\centering
\caption{Definitions of h\'a\v cek coefficients.}
\setlength{\tabcolsep}{5pt}
\renewcommand{\arraystretch}{1.5}
\begin{tabular}{ccc}
\hline\hline
H\'a\v cek coefficient& Combination & Number \\
\hline
&&\\[-10pt]
$ \hb{}^*_J \equiv \hb{}^+_J $&$
\fr{1}{\ga}( b_J-m_\mu g^{(A)}_J) 
+ \fr{1}{2}\ep_{JKL} H_{KL} + m_\mu d_{JT} 
+ \fr{1}{2\ga}(1+\fr{3}{2}\be^2\ga^2)m_\mu
\ep_{JKL}g^{(M)}_{KLT}
$&$ 3 $\\
$ \hb{}_J \equiv -\hb{}^-_J $&$
\fr{1}{\ga}( b_J-m_\mu g^{(A)}_J) 
- \fr{1}{2}\ep_{JKL} H_{KL} - m_\mu d_{JT}
+ \fr{1}{2\ga}(1+\fr{3}{2}\be^2\ga^2)m_\mu
\ep_{JKL}g^{(M)}_{KLT}
$&$ 3 $\\
$ \hb_T $&$
\fr{1}{\ga}(b_T-m_\mu g^{(A)}_T)
+ \fr{1}{\ga}(1-\fr{3}{2}\be^2\ga^2)m_\mu g^{(M)}_{XYZ}
$&$ 1 $\\
$ \hg_T $&$
\fr{1}{\ga}(b_T-m_\mu g^{(A)}_T) 
- \fr{2}{\ga}(1+\fr{3}{2}\be^2\ga^2) m_\mu g^{(M)}_{XYZ}
$&$ 1 $\\
$ \hh_{TX} $&$
H_{TX}-m_\mu d_{ZY}
-\fr{1}{\ga}(1-\fr{3}{2}\be^2\ga^2)m_\mu ( g^{(M)}_{TXT} 
- g^{(M)}_{XYY})
$&$ $\\
$ \hh_{TY} $&$
H_{TY}-m_\mu d_{XZ}
-\fr{1}{\ga}(1-\fr{3}{2}\be^2\ga^2)m_\mu( g^{(M)}_{TYT} 
- g^{(M)}_{YZZ})
$&$ $\\
$ \hh_{TZ} $&$
H_{TZ}-m_\mu d_{YX}
-\fr{1}{\ga}(1-\fr{3}{2}\be^2\ga^2)m_\mu( g^{(M)}_{TZT} 
- g^{(M)}_{ZXX})
$&$ 3 $\\
$ \hd_\pm $&$
m_\mu (d_{XX} \pm d_{YY})
$&$ 2 $\\
$ \hd_Q $&$
m_\mu (d_{XX}+d_{YY}-2d_{ZZ}) 
+ \fr{3}{\ga}(1-\fr{1}{2}\be^2\ga^2)m_\mu g^{(M)}_{XYZ}
$&$ 1 $\\
$ \hd_{XY} $&$
m_\mu (d_{XY} + d_{YX}) 
- \fr{1}{\ga}(1+\fr{3}{2}\be^2\ga^2) m_\mu (g^{(M)}_{TZT} 
+ 2 g^{(M)}_{ZXX})
$&$ $\\
$ \hd_{YZ} $&$
m_\mu (d_{YZ} + d_{ZY})
- \fr{1}{\ga}(1+\fr{3}{2}\be^2\ga^2) m_\mu (g^{(M)}_{TXT} 
+ 2 g^{(M)}_{XYY})
$&$ $\\
$ \hd_{ZX} $&$
m_\mu (d_{ZX}+ d_{XZ}) 
- \fr{1}{\ga}(1+\fr{3}{2}\be^2\ga^2) m_\mu (g^{(M)}_{TYT} 
+ 2 g^{(M)}_{YZZ})
$&$ 3 $\\
$ \hg_c $&$
\fr{1}{\ga} m_\mu (2g^{(M)}_{XYZ} + g^{(M)}_{YZX}
- \frac{3}{2}\be^2\ga^2 g^{(M)}_{ZXY})
$&$ 1 $\\
$ \hg_Q $&$
- \fr{3}{\ga}m_\mu (g^{(M)}_{TXX}+g^{(M)}_{TYY})
$&$ 1 $\\
$ \hg_{XZ} $&$
-\fr{1}{\ga}(1+\fr{3}{2}\be^2\ga^2)m_\mu (2g^{(M)}_{TXT}+g^{(M)}_{XYY})
$&$ $\\
$ \hg_{XY} $&$
-\fr{1}{\ga}(1+\fr{3}{2}\be^2\ga^2)m_\mu(g^{(M)}_{TXT}-g^{(M)}_{XYY})
$&$ $\\
$ \hg_{YX} $&$
-\fr{1}{\ga}(1+\fr{3}{2}\be^2\ga^2)m_\mu(2g^{(M)}_{TYT}+g^{(M)}_{YZZ})
$&$ $\\
$ \hg_{YZ} $&$
-\fr{1}{\ga}(1+\fr{3}{2}\be^2\ga^2)m_\mu(g^{(M)}_{TYT}-g^{(M)}_{YZZ})
$&$ $\\
$ \hg_{ZY} $&$
-\fr{1}{\ga}(1+\fr{3}{2}\be^2\ga^2)m_\mu(2g^{(M)}_{TZT}+g^{(M)}_{ZXX})
$&$ $\\
$ \hg_{ZX} $&$
-\fr{1}{\ga}(1+\fr{3}{2}\be^2\ga^2)m_\mu(g^{(M)}_{TZT}-g^{(M)}_{ZXX})
$&$ 6 $\\
& & \hskip -30pt Total: 25 \\
\hline\hline
\end{tabular}
\label{hacekcoeff}
\end{table*}

In the presence of Lorentz violation,
the motion of the Earth about the Sun
can introduce distinct time variations in the anomaly frequencies,
offering an opportunity to gain sensitivity
to additional coefficients.
Comparatively few experimental studies
have been performed that take advantage 
of the changes in the Earth's boost
over the course of the solar year,
in part due to factors such as 
the extended period of data collection,
the necessary long-term stability of the apparatus,
and the statistical power required.
Recent analyses accounting in detail for boost effects
include ones performed 
with a dual Xe-He maser 
\cite{cane}
and using a spin-torsion pendulum 
\cite{heckel}.
An analogous investigation is feasible
for the muon anomaly frequency,
with the added bonus that 
boost effects for both the muon and the antimuon
can be studied,
at least in principle.

In this subsection,
we consider boost signals
arising from minimal-SME operators in the muon sector
at leading relativistic order.
The spherical decomposition is well suited
for analyses of rotational properties
but is cumbersome for boosts,
so we work instead with cartesian coefficients for Lorentz violation.
The nonminimal cartesian coefficients could also be studied,
but the corresponding analysis
is more involved and lies outside our present scope.
The analysis here shows that measurements of the anomaly frequency
at existing and planned precisions
can yield sensitivities at the Planck-suppressed level
to 25 of the 44 independent observables
for Lorentz violation in the minimal-SME muon sector.
Most of these are unmeasured to date.

In standard coordinates
in the laboratory frame 
\cite{sunframe}, 
the correction to the anomaly frequency due to Lorentz violation is 
\bea
\de\om_a^\pm &=& \pm 2\hb{}^\pm_3
\nonumber\\
&\equiv&
\pm \frac 1\ga ( b_3 - m_\mu g^{({\rm A})}_3 )
+ m_\mu d_{30} + H_{12}
\nonumber\\
&& 
\pm \frac 1 \ga (1+\frac 32\be^2 \ga^2) m_\mu g^{({\rm M})}_{120} .
\label{boostb}
\eea
As discussed in Sec.\ \ref{Sidereal variations},
expressing the coefficients in the Sun-centered frame
reveals that the Earth's rotation 
introduces dependence on the coefficient combinations 
$\hb{}^\pm_X$, $\hb{}^\pm_Y$
given in Eq.\ \rf{bhacek}.
The Earth's boost $\be_\oplus \simeq 10^{-4}$ about the Sun 
produces sensitivity to additional coefficient combinations,
as does the laboratory boost $\be_L \simeq 10^{-6}$
due to the surface velocity from the Earth's rotation. 
Although the sensitivities to additional coefficients
are comparatively suppressed by the boost factors,
the experimental precision nonetheless would suffice
to yield results of definite interest.

The relativistic corrections to the anomaly frequencies
at leading order in $\be_\oplus$ and $\be_L$ can be obtained 
by transforming from the Sun-centered frame to the laboratory frame.
The transformation can be separated into two steps
\cite{sunframe}:
an instantaneous boost from the Sun-centered frame 
to a nonrotating frame at the Earth's surface,
followed by a rotation to the laboratory frame. 
The resulting expressions for $\hb{}^\pm_3$
in terms of coefficients in the Sun-centered frame
are given in Table \ref{orbitalmuon}.
In this table,
we denote
the Earth sidereal rotation frequency by 
$\om_\oplus \simeq 2\pi/(23 {\rm ~h} ~56 {\rm ~m})$ as before,
the Earth orbital frequency by 
$\Om_{\oplus} \simeq 2\pi/(365.26 {\rm ~d})$,
the Earth orbital tilt by $\et\simeq 23.5^\circ$,
and the colatitude of the laboratory by $\ch$.
The explicit expressions for $\hb{}^\pm_3$
are obtained for each particle
by multiplying all the factors in a particular row
and adding the contributions from all rows.
The coefficient factors that appear are expressed
in terms of h\'a\v cek coefficients,
which are convenient combinations of the basic cartesian coefficients
chosen to reduce to the standard tilde combinations
\cite{tables} 
in the nonrelativistic limit.
Explicit expressions for the h\'a\v cek coefficients
in terms of cartesian coefficients are given in Table \ref{hacekcoeff}.

The above discussion shows that
one goal of a search for Lorentz and CPT violation
using data from $g-2$ experiments
is to report sensitivities to the combinations 
of h\'a\v cek coefficients appearing in Table \ref{orbitalmuon}.
The terms independent of sidereal time
can be studied by comparing anomaly frequencies,
as described in Sec.\ \ref{Muon-antimuon comparison},
while those depending on $\codt$ and $\sodt$
but not on $\cto$ or $\sto$ 
lead to constraints
on $\hb{}^\pm_X$, $\hb{}^\pm_Y$ using the Earth's rotation,
as in Sec.\ \ref{Sidereal variations}.
All other terms represent effects due to boosts.
Each anomaly frequency acquires two contributions
from the Earth's orbital motion
depending on time as $\cto$ or $\sto$.
Other terms involving the Earth's boost
vary as products of $\codt$ or $\sodt$ with $\cto$ or $\sto$ 
and hence oscillate predominantly at the sidereal frequency
with a slow Earth-orbital variation superposed.
The remaining terms are suppressed
by the laboratory boost $\be_L$
and are either constant or vary with sidereal time.
Note that some of the latter oscillate at twice the sidereal frequency.

As before,
further insight can be gained by considering
bounds on individual h\'a\v cek coefficients 
assuming all others vanish.
In this scenario,
terms suppressed by $\be_L$ can reasonably be neglected in the analysis
because they are suppressed by a factor of 100 or more 
relative to all the others 
and because all individual coefficients appearing in these terms 
are also present elsewhere in the expressions
for the anomaly frequencies.
It therefore suffices to analyze 
experimental data for either the muon or the antimuon 
to obtain six independent results proportional to $\be_\oplus$,
corresponding to the six different component oscillations 
involving $\om_\oplus$ and $\Om_\oplus$.
Assuming sufficient statistical power in the data
and the design reach of order 0.1 ppm
for the forthcoming Fermilab
\cite{fermilab}
and J-PARC
\cite{j-parc}
experiments,
it appears plausible that sensitivities 
of order $10^{-20}$ GeV or better 
could be attained for each of six measurements 
on the antimuon anomaly frequency
involving the Earth boost $\be_\oplus$.
If muons are also available,
another six independent constraints can be obtained.
Taking one coefficient at a time,
these measurements would yield Planck-scale sensitivity 
to 25 of the 44 observables for muons in the minimal SME.
These 25 observables can be taken as the 25 h\'a\v cek
coefficients provided in the first column of Table \ref{hacekcoeff},
or equivalently as the 25 independent cartesian coefficients
appearing in the combinations listed in the second column.

\subsection{The apparent anomaly discrepancy}
\label{The anomaly discrepancy}

Calculations of the muon anomaly $a \equiv (g - 2)/2$
performed in the context of the SM 
\cite{jn}
produce a result lying about three standard deviations below the value 
measured by the BNL experiment
\cite{bnl,blum}.
The apparent discrepancy 
$\De a \equiv a_{\rm expt} - a_{\rm SM}$
could originate 
from comparatively prosaic sources
such as a statistical fluctuation in the experiment
or uncertainties in the SM theory,
or more dramatically from new physics beyond the SM.
Typical one-loop corrections arising from Lorentz-invariant new physics 
with coupling $g$ and mass scale $M$
contribute at order $g^2(m_\mu/M)^2$,
leading to a variety of predicted signals 
in existing experiments.
As one example,
the apparent anomaly discrepancy 
can be reproduced in unified models with
vector-like leptons having couplings $g\simeq 1/2$
and masses $\simeq 150$ GeV,
yielding concomitant signals at the LHC 
\cite{drs}.
Calculations of one-loop corrections to the anomaly
in special Lorentz-violating models
have also been performed
\cite{lvg-2}.

Here,
we consider a different idea,
based on the result \rf{omega-sun-frame}
showing that the presence of Lorentz violation
can shift the measured value of $\om_a^\pm$.
An appropriate shift of this type could induce an observed discrepancy
in the inferred value of the anomaly.
Indeed,
the apparent discrepancy $\De a$ would be reproduced
by a shift in the anomaly frequency of 
$\De \om_a \simeq 2\times 10^{-24}$ GeV.
It is then natural to ask 
whether any coefficients exist that can achieve this shift
while remaining compatible with existing constraints and,
if so,
what predictions this might yield for future experiments. 

Since appropriate coefficients for this purpose
must of necessity affect the anomaly frequency,
inspection of Eq.\ \rf{omega-sun-frame}
reveals that they must be a subset of 
$\hg^{(d)}_{njm}$ and $\hh^{(d)}_{njm}$. 
However,
the BNL data offer no indication that the discrepancy $\De a$
differs significantly between muons and antimuons
\cite{bnl},
so it is reasonable to consider only CPT-even effects.
This limits attention to the coefficients $\hh^{(d)}_{njm}$.
It also has the advantage of bypassing the existing constraints
on $\hg^{(d)}_{njm}$ 
obtained from direct comparisons of $\om_a^\pm$
and listed in Table \ref{table:muonantimuon2}.
In addition,
the Lomb spectrum and power distribution
shown in Fig.\ 2 of Ref.\ \cite{bnl08}
are consistent with no sidereal signal in the anomaly frequency,
which suggests restricting attention
to the coefficients $\hh^{(d)}_{nj0}$ with $m=0$.

The simplest terms for CPT-even effects without sidereal variations 
are associated with isotropic Lorentz violation, $j=m=0$,
and for $d\leq 6$
only one $H$-type coefficient of this kind exists
\cite{km13}.
However,
as described in Sec.\ \ref{Theory},
nonzero contributions to the anomaly frequency
appear only for coefficients with odd $j$,
so purely isotropic terms cannot reproduce the apparent discrepancy.
Instead,
the available coefficients satisfying the above criteria with $d\leq 6$
turn out to include one with $d=3$, 
$\hh^{(3)}_{010}$,
and three with $d=5$,
$\hh^{(5)}_{010}$, 
$\hh^{(5)}_{210}$, and 
$\hh^{(5)}_{230}$.

Assuming only one coefficient is nonzero at a time,
we find the approximate values needed to generate the required shift 
$\De\om_a$ in the anomaly frequency are 
\bea
\hh^{(3)}_{010} &\simeq& 3 \times 10^{-24} {\rm ~GeV},
\nonumber\\
\hh^{(5)}_{010} &\simeq& \hh^{(5)}_{210}
\simeq \hh^{(5)}_{230} \simeq 3 \times 10^{-25} {\rm ~GeV}^{-1}.
\label{values}
\eea
Any one of these four values therefore suffices 
to reproduce the discrepancy $\De a$,
while somewhat smaller values are required
if more than one coefficient is nonzero. 

Some experimental bounds already exist on these coefficients,
obtained from comparisons of anomaly-frequency measurements
at the differing colatitudes of BNL and CERN
and presented in Table \ref{hcoeffresults}.
A related constraint on
$H_{XY} \equiv \sqrt{3/4\pi}~ \hh^{(3)}_{010}$
is reported in Ref.\ \cite{bnl08}.
All these limits are compatible
with any of the four nonzero values \rf{values}
needed to reproduce the discrepancy $\De a$. 
Moreover,
no other relevant constraints exist from
\mm\ spectroscopy or astrophysical observations.
As described in Sec.\ \ref{Muonic bound states},
limits from \mm\ hyperfine transitions involve sidereal variations
and hence involve contributions only from coefficients with $m\neq 0$,
while other \mm\ spectroscopy lacks sufficient sensitivity.
Also, 
at present astrophysical limits have been placed 
only on isotropic coefficients,
and no sensitivity to $H$-type coefficients has been identified
\cite{km13}.

Overall,
the nonzero values \rf{values} appear largely acceptable 
on theoretical grounds as well.
They are sufficiently small to be plausible
as Planck-suppressed contributions from an underlying theory.
For example,
the required value of $\hh^{(3)}_{010}$
is more than two orders of magnitude below the ratio 
$m_\mu^2/M_P\simeq 9.2 \times 10^{-22}$ GeV.
Also,
CPT-even Lorentz-violating operators 
arise naturally in some frameworks.
For example,
noncommutative quantum field theories
\cite{szabo}
intrinsically involve Lorentz violation
because the commutator of coordinates in the spacetime manifold
introduces an antisymmetric two-index object $\th^{\mu\nu}$
that provides an orientation to spacetime in a given inertial frame,
and in realistic models this generates naturally 
a subset of CPT-even Lorentz-violating operators in the SME
\cite{chklo}.

One open theoretical issue beyond our present scope 
concerns radiative corrections,
which could reasonably be expected to mix these coefficients with others 
and perhaps contribute to CPT-even Lorentz-violating effects in other species. 
Most and possibly all such effects can be expected to lie
beyond current sensitivities,
but a complete investigation of this would be of definite interest.
We also note a potential philosophical disadvantage
to the choice \rf{values}:
the absence of sidereal effects arises because all four coefficients 
are aligned relative to the $Z$ axis in the Sun-centered frame,
which implies the low-probability scenario 
that the effects producing the anomaly discrepancy 
are aligned with the Earth's rotation axis.
In a realistic model,
at least some nonzero off-axis components might be expected
in addition to the values \rf{values},
in which case the sidereal constraints of Table \ref{sidconst}
would come into play.
These additional components could plausibly come 
with trigonometric factors of order 0.1,
in which case any of the choices \rf{values} would remain viable.
Nonetheless,
it is reasonable to suppose that 
if Lorentz violation is indeed the origin of the anomaly discrepancy,
then sidereal signals can be expected near the present limits.

One distinctive prediction of the choices \rf{values}
is a variation of the shift $\De\om_a$ with the experimental colatitude.
Applying Eq.\ \rf{omega-sun-frame}
and using Table \ref{gjmvalues} for the relevant $G_{j0}$ values,
the model with nonzero $\hh^{(3)}_{010}$
can be expected to shift the anomaly frequency 
measured in the forthcoming Fermilab 
\cite{fermilab}
and J-PARC 
\cite{j-parc}
experiments
away from the SM prediction by 
$\De\om_a = 2 G_{10}(\ch) \hh^{(3)}_{010} 
\simeq 2\times 10^{-24}$ GeV $\simeq 0.5$ rad Hz,
with the predicted J-PARC value being about 10\% smaller
due to the differing colatitudes and $G_{10}$ values.
For any of the three $d=5$ coefficients,
we find 
$\De\om_a = 2 E_0^2 G_{j0}(\ch) \hh^{(5)}_{nj0}
\simeq 2 \times 10^{-24}$ GeV
$\simeq 0.5$ rad Hz
at Fermilab again,
but due primarily to the lower antimuon energy
the J-PARC value is predicted to be about 100 times smaller
for $j=1$ and about 70 times smaller for $j=3$.
Observation of this effect would represent a striking signal
in favor of these models.

As a final remark,
we note that it may seem tempting to try to relate
the muon anomaly discrepancy to the proton radius puzzle.
However,
in the context of Lorentz violation 
this appears difficult to achieve at best.
As described in Sec.\ \ref{Proton radius puzzle},
the proton radius puzzle represents
a comparatively large low-energy effect 
of order $\De E_{\rm Lamb} \simeq 3\times 10^{-13}$ GeV, 
while the anomaly discrepancy is a much smaller high-energy effect 
of order $\De \om_a \simeq 2\times 10^{-24}$ GeV.
Although nonminimal Lorentz violation 
can naturally introduce an energy dependence,
the corresponding effects typically grow with energy rather than decreasing.
It therefore appears challenging 
to reproduce both observed phenomena with a single SME coefficient,
even without considering more detailed issues such as the spin dependence
of the effects.

\section{Summary and discussion}
\label{Summary}

This work has explored some prospects
for using laboratory experiments with muons and antimuons 
to search for Lorentz and CPT violation.
The first part of the paper concerns
spectroscopic measurements on muonic bound states.
Following a discussion of general features in Sec.\ \ref{Basics},
we begin by considering \mm\ transitions in Sec.\ \ref{Muonium}. 
Signals of Lorentz and CPT violation in \mm\ hyperfine transitions
are given by Eq.\ \rf{Mhyper},
and using published experimental results
we compile constraints on various nonrelativistic
and spherical coefficients in 
Tables \ref{table: NRHyper} and \ref{table: Hyper}.
We next consider the $1S$-$2S$ transition and the Lamb shift in \mm.
These offer interesting options for exploring 
isotropic Lorentz and CPT violation,
and by comparing experimental and theoretical values
we extract the constraints on isotropic nonrelativistic coefficients
shown in Table \ref{table3}.

In Sec.\ \ref{Muonic hydrogen},
we turn to an investigation of the spectroscopy 
of muonic atoms and ions.
Following some general considerations,
we begin by examining possible future searches
using sidereal variations in \hm\ Zeeman transitions.
The frequency shift 
of the $2S^{F-1}_{1/2}$-$2P^F_{3/2}$ transitions
induced by Lorentz violation is given by Eq.\ \rf{wm1},
and our analysis shows that interesting sensitivities in future experiments
can be achieved.
Next, 
we consider the hypothesis that Lorentz violation
could be the origin of the proton radius puzzle,
which arises from an apparent disagreement
in the value of the proton charge radius
obtained from \hm\ spectroscopy and from other experiments.
Nonzero SME coefficients obeying Eq.\ \rf{Br}
would generate a frequency shift matching the observed effect
while remaining consistent with existing constraints.
We then turn to the issue of searching for Lorentz and CPT violation
when the Zeeman transitions are unresolved.
A method is proposed to constrain possible effects
by using the apparent broadening of the spectral lines
resulting from the breaking of rotational symmetry.
Finally,
the prospects are investigated for studying Lorentz and CPT violation 
using other muonic atoms and ions including
\dm, \tm, \hetm, \hefm, \lixm, \linm, \bem, and \bom.
The expression \rf{genres} governs the frequency shifts
in all these systems,
and Table \ref{LF} provides a comparative measure 
of the attainable sensitivities.

Section \ref{Muon magnetic moment} of this work
focuses on Lorentz and CPT tests
using measurements of the anomalous magnetic moments
of the muon and antimuon.
We begin in Sec.\ \ref{Theory}
with some basic theory,
which shows that the observable shifts 
of the anomaly frequencies $\om_a^\pm$ 
of the muon and antimuon are given by Eq.\ \rf{omega-sun-frame}.
Several methods are available to place interesting constraints
from existing and future data.
We first consider comparisons of the muon and antimuon 
anomaly frequencies,
using different schemes to separate constraints
on CPT-odd and CPT-even effects.
Existing data are used to extract limits
on various spherical coefficients,
including numerous first bounds on nonminimal operators.
The results are compiled in 
Tables \ref{table:muonantimuon2} and Table \ref{hcoeffresults}.

Next,
we address the information available in the time domain.
Sidereal variations are considered
in Sec.\ \ref{Sidereal variations}.
Existing data are used to place a variety of limits,
which are tabulated in Table \ref{sidconst}.
We then investigate signals associated with the Earth's changing boost
as it revolves about the Sun.
The resulting modulations in the anomaly frequency
include harmonics with both annual and sidereal periodicities,
which are gathered in Table \ref{orbitalmuon}.
Estimates for the attainable sensitivities to Lorentz violation
from studies of annual variations are given.
Finally,
we consider the prospects of accounting for the anomaly discrepancy
between existing experimental data and SM calculations
using Lorentz violation.
This would require nonzero coefficients,
as shown in Eq.\ \rf{values},
that are compatible with present constraints
and lead to striking predictions for signals
in forthcoming experiments.

Except for partial overlap with published results for the minimal SME,
the constraints displayed in the various tables in this work
represent first limits 
on the dominant effects of muon-sector Lorentz and CPT violation.
Many of the constraints achieved lie at or beyond
the level that might be expected from Planck-suppressed effects,
and numerous interesting options remain open
for further experimental study along the lines suggested here.
We note in passing that a comparable phenomenological treatment
of the nonminimal sectors for electrons and other first-generation particles 
is lacking in the literature to date.
This means,
for example, 
that the results in the present work
are currently the best available constraints
on nonminimal Lorentz and CPT violation for charged leptons. 

Despite the substantial broadening
of the scope of tests of Lorentz symmetry with muons
presented in this work,
the techniques presented span 
only a comparatively small fraction 
of the theoretically available possibilities for Lorentz violation.
Considerable room remains for investigation,
including both uncovering additional methods 
to measure effects from unconstrained terms in the kinematic Lagrange density
and also developing tools to study Lorentz-violating interactions with muons.
Possibilities along the latter lines include,
for example,  
studying the effects of minimal and nonminimal Lorentz violation 
on various muon decays,
which in general are affected at the level of both muon kinematics
and muon interactions
\cite{dkl}.
More extensive studies of muon propagation and interactions
may be feasible if a muon collider is eventually realized,
perhaps to serve as a factory for Higgs bosons
\cite{muoncollider}.

Gravitational interactions of muons also offer
an intriguing avenue for exploration.
The gravitational sector of the SME
includes Lorentz-violating muon couplings
with a variety of signals  
that are in principle accessible to experiment
\cite{akgrav}.
For example,
the old issue of whether antiparticles 
can gravitate differently from particles
\cite{nietogoldman}
can be directly approached 
using the general matter-gravity couplings in the SME framework
\cite{akjt}.
An experiment has been proposed 
to address this question for muons using \mm\ interferometry
\cite{kirch},
and \hm\ interferometry may also be an option
\cite{lesche}.

In the context of the minimal SME,
the signals for the \mm-interferometry experiment
are considered in Sec.\ IX C of Ref.\ \cite{akjt}.
The gravitational acceleration of \mm\ is affected
differently from that of other matter
and also has a component varying with time
as the Earth revolves about the Sun.
The former effect,
which can be understood as a violation of the weak equivalence principle
induced by Lorentz violation,
is the most natural candidate signal
for \mm\ interferometry.
A detailed investigation including also 
nonminimal gravitational couplings of the muon
is infeasible at present,
but we can use dimensional arguments 
to estimate the attainable sensitivity
to the corresponding coefficients for Lorentz violation.
The phase shift $\de\ph$ in the \mm\ interferometer
takes the form 
$\de\ph \approx \ph_0m_\mu^{d-4} \K^{(d)}$,
where $\K^{(d)}$ is a generic coefficient
controlling a Lorentz-violating operator of mass dimension $d$
in the muon-gravity sector.
The muon mass $m_\mu$ enters because
the proposed experiment would use nonrelativistic \mm,
so the relevant energy is effectively the muon mass.
The phase $\ph_0 = 2\pi g \ta^2/d$
depends on the gravitational acceleration $g$,
the time of flight $\ta$,
and the grating separation $d$.  
Assuming the \mm\ experiment achieves a precision of 10\%,
then we can estimate sensitivities to $\K^{(d)}$ 
of order 
$|\K^{(d)}| \lsim 10^ {d-5}$ GeV$^{4-d}$.
Note that in general we can expect accompanying
sidereal and annual signals as well.

Another promising subject awaiting careful investigation
is flavor-changing effects involving muons,
which are natural in the SME context
\cite{ck}.
Planned experiments searching for decays such as $\mu^\pm\to e^\pm \ga$,
which are forbidden in the SM 
but for which Lorentz-violating operators appear in the SME,
are projected to attain sensitivities of a few parts in $10^{17}$ 
\cite{muegamma}
and hence could be of interest in the context of Planck-suppressed signals.
The flavor-changing operators in the SME also predict signals 
in searches for \mm-\mmb\ oscillations,
for which the current sensitivity
lies at the level of parts in $10^{11}$
\cite{willmann}.
Although a comprehensive treatment of nonminimal interactions
is unavailable to date,
dominant effects in flavor-changing processes
may well appear in the kinematics,
for which general tools are in hand
\cite{km13}.
Evidently,
the unexplored territory in the muon sector remains large,
and there is considerable promise for future discovery
in a wide variety of experiments.

\section*{Acknowledgments}
\bigskip

This work was supported in part
by the U.S.\ Department of Energy
under grant DE-FG02-13ER42002,
by the Indiana University Center for Spacetime Symmetries,
and by 
the Conselho Nacional de Desenvolvimento Cientifico e Tecnol\'ogico
in Brazil.


\begin{thebibliography}{99}

\bibitem{an}
P.\ Kunze, Z.\ Phys.\ {\bf 83}, 1 (1933);
C.D.\ Anderson and S.H.\ Neddermeyer,
Phys.\ Rev.\ {\bf 50}, 263 (1936).

\bibitem{rh}
B.\ Rossi and D.B.\ Hall,
Phys.\ Rev.\ {\bf 59}, 223 (1941).

\bibitem{bailey} 
H.\ Bailey \etal,
Nature {\bf 268}, 301 (1977).

\bibitem{ksp}
V.A.\ Kosteleck\'y and S.\ Samuel,
Phys.\ Rev.\ D {\bf 39}, 683 (1989);
V.A.\ Kosteleck\'y and R.\ Potting,
Nucl.\ Phys.\ B {\bf 359}, 545 (1991);
Phys.\ Rev.\ D {\bf 51}, 3923 (1995). 

\bibitem{tables}
{\it Data Tables for Lorentz and CPT Violation, 2014 edition,}
V.A.\ Kosteleck\'y and N.\ Russell,
arXiv:0801.0287v7.

\bibitem{ck}
D.\ Colladay and V.A.\ Kosteleck\'y,
Phys.\ Rev.\ D {\bf 55}, 6760 (1997);
Phys.\ Rev.\ D {\bf 58}, 116002 (1998).

\bibitem{akgrav}
V.A.\ Kosteleck\'y,
Phys.\ Rev.\ D {\bf 69}, 105009 (2004).

\bibitem{owg}
O.W.\ Greenberg,
Phys.\ Rev.\ Lett.\ {\bf 89}, 231602 (2002).

\bibitem{bkl}
R.\ Bluhm, V.A.\ Kosteleck\'y, and C.D.\ Lane,
Phys.\ Rev.\ Lett.\ {\bf 84}, 1098 (2000).

\bibitem{hughes01}
V.W.\ Hughes \etal,
Phys.\ Rev.\ Lett.\ {\bf 87}, 111804 (2001).

\bibitem{bnl01}
M.\ Deile \etal,
hep-ex/0110044.

\bibitem{bnl08}
G.W.\ Bennett \etal,
Phys.\ Rev.\ Lett.\ {\bf 100}, 091602 (2008).

\bibitem{srf}
Y.V.\ Stadnik, B.M.\ Roberts, and V.V.\ Flambaum,
Phys.\ Rev.\ D {\bf 90}, 045035 (2014).

\bibitem{noordmans}
J.P.\ Noordmans,
{\it Lorentz Violation in the Weak Interaction},
Rijksuniversiteit Groningen thesis, 2014.

\bibitem{ba07}
B.\ Altschul,
Astropart.\ Phys.\ {\bf 28}, 380 (2007).

\bibitem{gm04}
O.\ Gagnon and G.D.\ Moore,
Phys.\ Rev.\ D {\bf 70}, 065002 (2004).

\bibitem{km13}
V.A.\ Kosteleck\'y and M.\ Mewes,
Phys.\ Rev.\ D {\bf 88}, 096006 (2013).

\bibitem{antognini}
A.\ Antognini \etal, 
Science {\bf 339}, 417 (2013);  
R.\ Pohl \etal, Nature {\bf 466}, 213 (2010).

\bibitem{deuterium}
R.\ Pohl \etal,
J.\ Phys.\ Conf.\ Ser.\ {\bf 264}, 012008 (2011).

\bibitem{crema}
F.\ Biraben \etal,
{\it Lamb Shift in Muonic Helium},
PSI proposal R10-01, January 2010;
A.\ Antognini \etal,
Can.\ J.\ Phys.\ {\bf 89}, 47 (2011).

\bibitem{fujiwara}
M.C.\ Fujiwara \etal,
Hyperf.\ Int.\ {\bf 118}, 151 (1999).

\bibitem{drakebyer}
G.W.F.\ Drake and L.L.\ Byer,
Phys.\ Rev.\ A {\bf 32}, 713 (1985).

\bibitem{bl}
S.J.\ Brodsky and R.F.\ Lebed,
Phys.\ Rev.\ Lett.\ {\bf 102}, 213401 (2009).

\bibitem{hps}
A.\ Banburski and P.\ Schuster,
Phys.\ Rev.\ D {\bf 86}, 093007 (2012).

\bibitem{jaw}
J.A.\ Wheeler,
Rev.\ Mod.\ Phys.\ {\bf 21}, 133 (1949).

\bibitem{eides}
M.I.\ Eides, H.\ Grotch, and V.A.\ Shelyuto, 
{\it Theory of Light Hydrogenic Bound States},
Springer, Berlin, 2007.

\bibitem{akrl}
V.A.\ Kosteleck\'y and R.\ Lehnert,
Phys.\ Rev.\ D {\bf 63}, 065008 (2001).

\bibitem{kla}
V.A.\ Kosteleck\'y and C.D.\ Lane,
Phys.\ Rev.\ D {\bf 60}, 116010 (1999);
J.\ Math.\ Phys.\ {\bf 40}, 6245 (1999).

\bibitem{km09}
V.A.\ Kosteleck\'y and M.\ Mewes,
Phys.\ Rev.\ D {\bf 80}, 015020 (2009).

\bibitem{sunframe}
R.\ Bluhm {\it et al.},
Phys.\ Rev.\ D {\bf 68}, 125008 (2003);
Phys.\ Rev.\ Lett.\ {\bf 88}, 090801 (2002);
V.A.\ Kosteleck\'y and M.\ Mewes,
Phys.\ Rev.\ D {\bf 66}, 056005 (2002).

\bibitem{ak98}
V.A.\ Kosteleck\'y,
Phys.\ Rev.\ Lett.\ {\bf 80}, 1818 (1998).

\bibitem{liu99}
W.\ Liu \etal,
Phys.\ Rev.\ Lett.\ {\bf 82}, 711 (1999).

\bibitem{we}
E.P.\ Wigner, Z.\ Physik {\bf 43}, 624 (1927);
C.\ Eckart, Rev.\ Mod.\ Phys.\ {\bf 2}, 305 (1930).

\bibitem{shimomura}
K.\ Shimomura, 
{\it Precision measurement of muonium hyperfine structure 
and muon magnetic moment},
proposal 2011MS01 (2011);
K.\ Sasaki \etal,
J.\ Phys.\ Conf.\ Ser.\ {\bf 408}, 012074 (2013).

\bibitem{hydrogen}
R.\ Bluhm \etal,
Phys.\ Rev.\ Lett.\ {\bf 82}, 2254 (1999);
B.\ Altschul,
Phys.\ Rev.\ D {\bf 81}, 041701 (2010);
A.\ Matveev \etal,
Phys.\ Rev.\ Lett.\ {\bf 110}, 230801 (2013).

\bibitem{meyer}
V.\ Meyer \etal, 
Phys.\ Rev.\ Lett.\ {\bf 84}, 1136 (2000).

\bibitem{mohr}
P.J.\ Mohr, B.N.\ Taylor, and D.B.\ Newell, 
Rev.\ Mod.\ Phys.\ {\bf 84}, 1527 (2012).

\bibitem{woodle}
K.A.\ Woodle \etal, 
Phys.\ Rev.\ A {\bf 41}, 93 (1990).

\bibitem{pohl}
For a recent review see,
for example,
R.\ Pohl, R.\ Gilman, G.A.\ Miller, and K.\ Pachucki,
Annu.\ Rev.\ Nucl.\ Part.\ Sci. {\bf 63}, 175 (2013).

\bibitem{wigner}
E.P.\ Wigner,
{\it Group Theory},
Academic, New York, 1959.

\bibitem{pcrth}
K. Pachucki,
Phys.\ Rev.\ A {\bf 60}, 3593 (1999);
U.D.\ Jentschura,
Ann.\ Phys.\ {\bf 326}, 500 (2011);
E.\ Borie,
Ann.\ Phys.\ {\bf 327}, 733 (2012);
S.G.\ Karshenboim, V.G.\ Ivanov, and E.Y.\ Korzinin,
Phys.\ Rev.\ A {\bf 85}, 032509 (2012).

\bibitem{antognini2}
A.\ Antognini \etal,
Ann.\ Phys.\ {\bf 331}, 127 (2013).

\bibitem{pzrth}
A.P.\ Martynenko, 
Phys.\ Rev.\ A {\bf 71}, 022506 (2005);
C.E.\ Carlson, V.\ Nazaryan, and K.\ Griffioen,
Phys.\ Rev.\ A {\bf 83}, 042509 (2011).

\bibitem{cern}
J.\ Bailey \etal,
Nucl.\ Phys.\ B {\bf 150}, 1 (1979).

\bibitem{bnl}
G.W.\ Bennett \etal, 
Phys.\ Rev.\ D {\bf 73}, 072003 (2006).

\bibitem{fermilab}
R.M.\ Carey \etal,
Fermilab proposal 0989 (2009).

\bibitem{j-parc}
M.\ Aoki \etal,
J-PARC proposal J-PARC-PAC2009-12 (2009).

\bibitem{finsler1}
V.A.\ Kosteleck\'y,
Phys.\ Lett.\ B {\bf 701}, 137 (2011).

\bibitem{finsler2}
M.\ Schreck, arXiv:1405.5518;
V.A.\ Kosteleck\'y, N.\ Russell, and R.\ Tso,
Phys.\ Lett.\ B {\bf 716}, 470 (2012);
D.\ Colladay and P.\ McDonald,
Phys.\ Rev.\ D {\bf 85}, 044042 (2012);
V.A.\ Kosteleck\'y and N.\ Russell,
Phys.\ Lett.\ B {\bf 693}, 443 (2010).
	
\bibitem{krt08}
V.A.\ Kosteleck\'y, N.\ Russell, and J.\ Tasson,
Phys.\ Rev.\ Lett.\ {\bf 100}, 111102 (2008).

\bibitem{fittante}
A.\ Fittante and N.\ Russell,
J.\ Phys.\ G {\bf 39}, 125004 (2012).

\bibitem{cane}
F.\ Can\`e \etal,
Phys.\ Rev.\ Lett.\ {\bf 93}, 230801 (2004).

\bibitem{heckel}
B.R.\ Heckel, E.G.\ Adelberger, C.E.\ Cramer, 
T.S.\ Cook, S.\ Schlamminger, and U.\ Schmidt,
Phys.\ Rev.\ D {\bf 78}, 092006 (2008).

\bibitem{jn}
For a review see, for example,
F.\ Jegerlehner and A.\ Nyffeler,
Phys.\ Rep.\ {\bf 477}, 1 (2009).

\bibitem{blum}
T.\ Blum \etal,
arXiv:1311.2198.

\bibitem{drs}
R.\ Derm\'i\v sek, A.\ Raval, and S.\ Shin,
Phys.\ Rev.\ D {\bf 90}, 034023 (2014).

\bibitem{lvg-2}
A.\ Moyotl, H.\ Novales-S\'anchez, J.J.\ Toscano, and E.S.\ Tututi,
Int.\ J.\ Mod.\ Phys.\ A {\bf 29}, 1450039 (2014);
Int.\ J.\ Mod.\ Phys.\ A {\bf 29}, 1450107 (2014);
C.D.\ Carone, M.\ Sher, and M.\ Vanderhaeghen,
Phys.\ Rev.\ D {\bf 74}, 077901 (2006);
W.F.\ Chen and G.\ Kunstatter,
Phys.\ Rev.\ D {\bf 62}, 105029 (2000).

\bibitem{szabo}
For a review see,
for example,
R.J.\ Szabo,
Phys.\ Rep.\ {\bf 378}, 207 (2003).

\bibitem{chklo}
S.M.\ Carroll {\it et al.},
Phys.\ Rev.\ Lett.\ {\bf 87}, 141601 (2001).

\bibitem{dkl}
J.S.\ D\'\i az, V.A.\ Kosteleck\'y, and R.\ Lehnert,
Phys.\ Rev.\ D {\bf 88}, 071902(R) (2013).

\bibitem{muoncollider}
See, for example, 
Y.\ Alexahin \etal,
arXiv:1308.2143.

\bibitem{nietogoldman}
See, for example,
M.M.\ Nieto and J.T.\ Goldman,
Phys.\ Rep.\ {\bf 205}, 221 (1991).

\bibitem{akjt}
V.A.\ Kosteleck\'y and J.D.\ Tasson,
Phys.\ Rev.\ Lett.\ {\bf 102}, 010402 (2009);
Phys.\ Rev.\ D {\bf 83}, 016013 (2011).

\bibitem{kirch}
K.\ Kirch,
arXiv:physics/0702143.

\bibitem{lesche}
B.\ Lesche,
Gen.\ Rel.\ Grav.\ {\bf 21}, 623 (1989).

\bibitem{muegamma}
See, for example,
D.\ Brown,
Nucl.\ Phys.\ B Proc.\ Suppl.\ {\bf 248}, 41 (2014);
A.\ Edmonds,
Nucl.\ Phys.\ B Proc.\ Suppl.\ {\bf 248}, 47 (2014).

\bibitem{willmann}
L.\ Willmann \etal,
Phys.\ Rev.\ Lett.\ {\bf 82}, 49 (1999).

\end{thebibliography}
\end{document}